\documentclass[fleqn,usenatbib]{mnras}

\usepackage{newtxtext,newtxmath}

\usepackage[T1]{fontenc}
\usepackage{xcolor} 
\usepackage[dvipsnames]{xcolor}

\DeclareRobustCommand{\VAN}[3]{#2}
\let\VANthebibliography\thebibliography
\def\thebibliography{\DeclareRobustCommand{\VAN}[3]{##3}\VANthebibliography}

\usepackage{graphicx}	
\usepackage{amsmath}	
\usepackage[normalem]{ulem} 
\usepackage{CJK}
\usepackage[version=4]{mhchem} 
\usepackage{natbib}
\usepackage{hyperref}

\newcommand{\massflux}{kg m$^{-2}$ s$^{-1}$}
\newcommand{\g}{\,m\,s$^{-2}$}
\newcommand{\Mtwentyfive}{\hyperlink{cite.Mak_etal2025}{M25}}

\title[Part I: Haze Size Distribution over Two Limbs]{Flow-Driven Limb-Asymmetry of Haze Distribution Part I: An Analytical Framework for Predicting the Size Distribution of Photochemical Hazes Across the Two Limbs of hot-Jupiters}

\author[M. T. Mak et al.]{Mei Ting Mak,$^{1,2}$\thanks{Croucher Postdoctoral Fellow}\thanks{E-mail: martha.mak@physics.ox.ac.uk}
Thaddeus D. Komacek$^{1}$,
Nathan J. Mayne$^{2}$,
David K. Sing$^{3,4}$
\\
$^{1}$Atmospheric, Oceanic, and Planetary Physics Department, University of Oxford, OX1 3PU, UK\\
$^{2}$Department of Physics and Astronomy, University of Exeter, Exeter EX4 4QL, UK\\
$^{3}$Department of Physics and Astronomy, Johns Hopkins University, Baltimore, MD 21218, USA\\
$^{4}$Department of Earth and Planetary Sciences, Johns Hopkins University, Baltimore, MD 21218, USA
}
\date{Accepted XXX. Received YYY; in original form ZZZ}

\pubyear{\the\year{}}

\begin{document}
\label{firstpage}
\pagerange{\pageref{firstpage}--\pageref{lastpage}}
\maketitle

\begin{abstract}
Photochemical haze, a common aerosol type expected to form in the atmospheres of hot-Jupiters, can become concentrated to different extents between the morning and evening limbs depending on the balance between advection, gravitational settling, and radiation pressure. We present a  analytical framework incorporating the effect of gravity, planetary radius, and stellar flux, alongside the particle size of the haze on its resulting relative distribution between the two limbs. Using this framework and further comparing with 3D climate simulations, our framework provides a reasonable first-order estimate of the maximum radius of haze particles which would reach the morning limb and subsequently be trapped by the nightside gyres, resulting in a higher or comparable concentration of haze over the morning limb compared to the evening limb for a given hot-Jupiter atmosphere. We find that the framework performs best for higher-gravity planets, where the transport of haze particles is more strongly controlled by gravitational settling and therefore less sensitive to the approximations made in describing the atmospheric circulation. We further show that for low-gravity hot-Jupiters, even large haze particles can be readily transported to the morning limb before being removed by gravitational settling, whereas for high-gravity hot-Jupiters only small particles can survive transport to the morning limb. Our novel framework provides a rapid way to understand the transport of haze and plan limb asymmetry observations with JWST, constraining the parameter space exploration for full-scale computationally expensive 3D simulations.
\end{abstract}




\section{Introduction}
\label{sec:introduction}
State-of-the-art telescopes, such as the James Webb Space Telescope (JWST), have enabled the characterisation of exoplanets via transmission spectroscopy with a precision down to a few tens to hundreds of parts per million (ppm), allowing spectral comparisons between the morning (leading/western -- 90$^{\circ}$ west of the substellar point) and evening (trailing/eastern -- 90$^{\circ}$ east of the substellar point) planetary limbs of hot-Jupiters \citep{Rustamkulov_etal2023,Carter_etal2024,Espinoza_2024,Murphy_etal2024,Mukherjee_etal2026}. Understanding the longitudinal differences in transmission spectra allows for the study of transport, production and removal of materials such as clouds and hazes within the atmosphere, thereby placing strong constraints on the atmospheric dynamics \citep{Kempton_etal2017,Powell_etal2019,Christie_etal2021,Tsai_etal2023,Savel_etal2023,Zamyatina_etal2024,Fu_etal2025,Mak_etal2025, Owen_etal2025}.

The presence of photochemical haze can be diagnosed by comparing limb asymmetries in transmission spectra. Photochemical hazes, the solid-state products from light-driven photochemistry \citep{Zahnle_etal2009a,Gao_etal2021}, are assumed to form on the day side of the planet (where it is receiving stellar flux), with a majority of the particles transported eastward due to the prograde super-rotating jet in the atmosphere \citep{Showman_and_Polvani_2011,Steinrueck_etal2021,Mak_etal2025}. During this process, a majority of particles will first pass through the evening limb and then the morning limb, while some of them are trapped over the mid-latitudes nightside gyres (close to the morning limb) that form due to the planetary scale standing Rossby-Kelvin wave pattern driven by the day-night irradiation contrast from tidal locking, similar to the transport of many other tracer species \citep{Steinrueck_etal2021,Tsai_etal2023,Braam_etal2023,Zamyatina_etal2024,Mak_etal2025}. 

These haze particles are subject to gravitational settling \citep{Parmentier_etal2013,Fu_etal2025}. \citet{Kempton_etal2017} suggested that as haze particles are transported eastward by the jet, they would settle to the deep atmosphere over the nightside before reaching the morning limb, creating a hazy spectrum over the evening limb and a clear spectrum over the morning limb. \citet{Steinrueck_etal2021} used a 3D General Circulation Model (GCM), SPARC/MITgcm, and illustrated this effect when simulating the presence of haze with chosen particle radii of 30\,nm, 100\,nm, 300\,nm and 1000\,nm in the atmosphere of HD\,189733b. Recent studies have also highlighted the impact of radiation pressure in altering the limb distribution of hazes, as the strong flux of stellar radiation received by hot-Jupiters can strengthen the gravitational settling effect of solid particles on the dayside \citep{Fu_etal2025,Owen_etal2025}.

\citet{Kempton_etal2017} also argued that efficient mixing from the prograde jet and mid-latitudinal retrograde advection can act to increase haze concentrations over the morning limb. Results from \citet{Steinrueck_etal2021} when assuming a haze particle radius of $<$30\,nm in HD\,189733b also show a higher haze concentration over the morning limb relative to the evening limb due to the weakened gravitational settling for small particles. Extending this to a larger sample of hot-Jupiters, \citet{Mak_etal2025} (\Mtwentyfive, hereafter) demonstrated a similar result for small haze particles (radius of 1.5\,nm) in HD\,189733b, HD\,209458b and WASP-39b. \Mtwentyfive\ showed an increase in the scale height over the morning limb due to the shortwave absorbing properties of haze, driving radiative heating within the atmosphere. \Mtwentyfive\ further demonstrated that, for WASP-39b, the morning limb could present a larger transit depth than the evening limb at short wavelengths due to both the increased haze opacity and the hotter thermal structure caused by radiative heating from haze, while both spectra still retain a strong scattering slope. This is in contrast to the usual case of a larger transit depth over the evening limb due to its higher temperature. Even though \Mtwentyfive\ did not find this reversed effect in simulations of HD\,189733b and HD\,209458b, they suggested that a larger transit depth of the morning limb at short wavelengths together with a steep scattering slope, if detected, could act as a diagnostic for the presence of haze.


Motivated by this observational diagnostic of haze as shown in WASP-39b, which can be explained by a higher concentration of haze over the morning limb, this work focuses on developing and introducing a physically motivated analytical framework that captures the first-order balance between advection, settling, and radiation pressure. This framework allows for a simple and computationally inexpensive estimation of the maximum radius required by haze particles to survive transport to the morning limb before gravitational settling dominates. The subsequent trapping of these particles within the nightside gyres, which can lead to a higher or comparable haze concentration over the morning limb than the evening limb, is expected as seen from previous 3D simulations (\citealt{Steinrueck_etal2021}, \Mtwentyfive, \citealt{Lee_etal2026}) but not explicitly modelled within the present analytical framework. This simple diagnostic can be readily applied to observational planning and interpretation across many hot-Jupiters prior to running full-scale, computationally-expensive 3D GCM simulations. We note that the purpose of this work is to provide an overarching framework and diagnostic, rather than a fully quantitative predictive model. A more detailed and comprehensive investigation, which focuses on explaining the asymmetrical haze distribution using the full dynamical behaviour within GCMs, as well as on the limitations of the current framework and the mechanisms that drive the deviation of the approximated maximum radius using our framework from GCM output, will be presented in a follow-up study (Mak et al. in prep). In Sec.~\ref{sec:method}, we introduce the mathematical tools used to describe the balance of the strength from advection, gravitational settling and radiation pressure, and the 3D GCM used in running simulations over a range of radii and mass flux of haze particles in the atmosphere of WASP-39b (setup similar to \Mtwentyfive). In Sec.~\ref{sec:results}, we present the simulation results for WASP-39b and compare them, along with literature on other hazy hot-Jupiters with different surface gravities, against the estimations based on the introduced framework. In Sec.~\ref{sec:discussions}, we explore the reliability and limitations of this framework, and we present a summary of this work in Sec.~\ref{sec:conclusions}.

\section{Method}
\label{sec:method}

\subsection{Analytical Theory}
\label{subsec:analytical_theory}
For the purposes of our analytical theory, we simplify the complex three-dimensional atmospheric circulation of hot-Jupiters to two dimensions, the horizontal and vertical. We then approximate haze-driven limb asymmetry through balancing the haze particle motion in these two directions. This treatment assumes a non-zero haze production on the dayside of the planet, and then determines the horizontal and vertical advection conditions, alongside the particle properties with relation to settling, to identify when a limb asymmetry in the haze particle size distribution is possible.  


\subsubsection{Horizontal Dimension}
\label{subsec:horizontal_dimension}
We focus on the equatorial region, as it is hotter and therefore has a larger scale height, causing transmission spectroscopy to be more sensitive to this region. Hence we assume for simplicity that the horizontal motion is dictated by the super-rotating jet. We follow \citet{Zhang_and_Showman_2017} and characterise this motion with the bulk root-mean-square of the zonally averaged flow across the equatorial region $U_\text{rms}$, assuming it yields comparable speed to the super-rotating jet. We define the horizontal advection timescale $\tau_a$ to be 
\begin{equation}
\label{eqn:ta}
\tau_a = \frac{{\pi}R_p}{U_\text{rms}}\quad,
\end{equation}
where $R_p$ is the planetary radius and $\pi{R_p}$ is the hemispheric distance, which is also the distance between the evening and morning limb. 

Rather than adopting a fixed $\tau_a$, we allow $\tau_a$ to vary with planetary surface gravity, instellation and rotation rate as they alter the radiative timescale and hence the circulation strength \citep{Komacek_and_Showman_2016,Zhang_and_Showman_2017}. While $U_\text{rms}$ can be obtained directly from running 3D GCM simulations adopting specific stellar and planetary parameters, here we opt for an analytical estimation of $U_\text{rms}$ so that $\tau_a$ has the appropriate scaling across the range of hot-Jupiters explored in this study. In this work, we follow the treatment from \citet{Zhang_and_Showman_2017} and summarise the key equations adopted from their work to estimate $U_\text{rms}$.

Based on the momentum, energy, and continuity equations, while also assuming that the atmosphere is in hydrostatic balance, \citet{Zhang_and_Showman_2017} estimate $U_\text{rms}$ as 
\begin{equation}
\label{eqn:U_rms}
U_\text{rms} \sim \frac{2\gamma}{\alpha + \sqrt{\alpha^2 + 4\gamma^2}} U_\text{eq}\quad,
\end{equation}
where $\alpha$ and $\gamma$ are dimensionless quantities and $U_\text{eq}$ is the cyclostrophic wind speed induced by an equilibrium day-night temperature difference $\Delta{T}_\text{eq}$. $U_\text{eq}$ can also be explained as the reference wind speed that arises from day-night pressure gradient if $\Delta{T}_\text{eq}$ is in radiative equilibrium and if the planetary Rossby number exceeds 1. $\alpha$, $\gamma$ and $U_\text{eq}$ are given as
\begin{equation}
\label{eqn:U_eq}
\begin{aligned}
\alpha = 1 + \frac{\left( \Omega + \frac{1}{\tau_\text{drag}} \right) \tau^2_\text{wave}}{\tau_\text{rad}\Delta\ln{p}} \quad, \\
\gamma = \frac{\tau^2_\text{wave}}{\tau_\text{rad}\tau_\text{adv,eq}\Delta\ln{p}} \quad, \\
U_\text{eq} = \sqrt{ \frac{R\Delta{T}_\text{eq}\Delta\ln{p}}{2\mu} } \quad,
\end{aligned}
\end{equation}
respectively, where $\Omega$ is the planetary rotation rate, and $\tau_\text{drag}$, $\tau_\text{wave}$ and $\tau_\text{rad}$ are the drag, wave propagation and radiative timescales, respectively. $R$ and $\mu$ are the molar specific gas constant and the mean molecular weight, respectively. $\tau_\text{adv,eq}$ is given as $L/U_\text{eq}$ where $L$ is the horizontal length scale, taken to be $R_p$. $\Delta\ln p$ is given as $\ln[p_\text{ref}/p_{\tau=1}]$ where $p_\text{ref}$ is a deep reference pressure where the day-night temperature difference is negligible, typically at the order of $\sim10^3$\,mbar according to previous GCMs work on hot-Jupiters \citep[e.g.,][]{Drummond_etal2020,Zamyatina_etal2023}. $p_{\tau=1}$ is the pressure of the observable region probed by transmission at which the atmospheric opacity reaches 1. In this work, we assume that the transmission spectra can probe atmospheric layers down to $\sim$10\,mbar, based on previous 3D GCM studies of haze on hot-Jupiter \citep[e.g.,][]{Steinrueck_etal2023}. $\tau_\text{drag}$ characterises the Rayleigh drag crudely representing the effects of turbulence and magneto-hydrodynamics \citep{Koll_and_Komacek_2018}. Since this work is considering slightly colder hot-Jupiters which are expected to have a relatively weaker drag force \citep{Menou_2012}, we have assumed $\tau_\text{drag}=\infty$ for simplicity. For $\Delta{T}_\text{eq}$, we follows the approximation from \citet{Zhang_2020} which assumes $\Delta{T}_\text{eq}\sim T_\text{eq}$ where $T_\text{eq}$ is the equilibrium temperature of the planet.


To simplify the calculation, we have assumed an isothermal atmosphere ($d\ln T/d\ln p =0$, where $T$ is the local temperature) in the region observable via transmission spectra, allowing $\tau_\text{wave}$ to be estimated as 
\begin{equation}
\label{eqn:tau_wave}
\tau_\text{wave} \sim \frac{L}{NH} = L \left[ \frac{RT}{\mu} \left(\frac{R}{c_p} - \frac{d\ln T}{d\ln p} \right) \right] ^{-1/2}= \frac{L}{R} \left( \frac{\mu c_p}{T} \right)^{1/2} \quad,
\end{equation}
where $N$ and $H$ are the Brunt-V\"ais\"al\"a frequency and scale height, respectively, and $c_p$ is the molar heat capacity. Additionally, assuming a blackbody emission, $\tau_\text{rad}$ can be estimated as \citep{Showman_and_Gillot_2002}
\begin{equation}
\label{eqn:tau_rad}
\tau_\text{rad} \sim \frac{p_{\tau=1}}{g} \frac{c_p}{4\sigma T_\text{eq}^3} \quad,
\end{equation}
where $\sigma$ is the Stefan-Boltzmann constant. $T_\text{eq}$ is chosen in Equation~\ref{eqn:tau_rad} to represent temperature of the observable region of the atmosphere. We note that the adopted values for $\tau_\text{drag}$, $p_\text{ref}$, $p_{\tau=1}$ and $\Delta{T}_\text{eq}$ are treated as representative values for this first-order framework. Furthermore, as demonstrated in \citet{Roth_etal2024}, since $p/g=\rho z$ (from hydrostatic equilibrium) and $\tau_\text{opac} = \kappa \rho z$ where $\tau_\text{opac}$ is the optical depth and $\kappa$ is the opacity, Equation~\ref{eqn:tau_rad} can also be expressed as $\tau_\text{rad} \sim c_p/(4\sigma\kappa T_\text{eq}^3)$ at $p=p_{\tau=1}$. Assuming a fixed atmospheric composition across hot-Jupiters with a representative opacity, $c_p$, $\mu$ and $\kappa$ can be treated as approximately constant to first order, resulting in $\tau_\text{rad}$ primarily depends on temperature through the scaling of $1/T_\text{eq}^3$. We therefore retain Equation~(\ref{eqn:tau_rad}) as a first-order estimate of the radiative timescale, while noting that a fully self-consistent treatment would require detailed radiative transfer calculations coupled to the atmospheric structure of each planet. The impact of these approximations is discussed in Sec.~\ref{sec:discussions}.


\subsubsection{Vertical Dimension}
\label{subsec:vertical_dimension}
The vertical advection timescale $\tau_w$ can be defined as
\begin{equation}
\label{eqn:tw}
\tau_w = \frac{H}{W} \quad,
\end{equation} 
where $H$ is the scale height and $W$ is the nightside hemispherically-averaged downward vertical velocity. Although localised regions of both upwelling and downwelling are present in hot-Jupiter atmospheres, previous 3D studies (e.g., \citealp[]{Mayne_etal2014b}, \citealp[]{Drummond_etal2018b}, \Mtwentyfive) show that the nightside atmosphere exhibits a net downward circulation when averaged over the nightside hemisphere. We therefore adopt the hemispherically-averaged nightside downward velocity as a representative measure of the net large-scale vertical transport. Consequently, we use this net downward transport to define a characteristic vertical transport timescale. The large-scale overturning circulation of hot-Jupiters is, however, yet to be inferred observationally, so we acknowledge that other large-scale patterns are possible, and would lead to a different result to that presented herein. We emphasize that this work explores the regime of large-scale vertical motions. Turbulent mixing by small-scale unresolved eddies and diffusive processes are not explicitly included in our large-scale framework. The scale height $H$ is given by
\begin{equation}
\label{eqn:H}
H = \frac{k_\mathrm{B}T}{mg} \quad,
\end{equation}
where $k_\mathrm{B}$ is the Boltzmann constant and $m$ is the molecular mass ($m=\mu/N_\text{A}$, where $N_\text{A}$ is the Avogadro's constant). For hot-Jupiters, we assume the atmosphere to be \ce{H2}-dominated, resulting in $m = 3.34\times10^{-27}$\,kg. 

Similar to Sec.~\ref{subsec:horizontal_dimension}, $W$ can be diagnosed from 3D simulations or estimated through analytical theories. In this work, we follow \citet{Tan_2022} who employ a scaling relation of ageostrophic flow $U_\text{ageo}$ to estimate vertical motions. Ageostrophic flow scales as 
\begin{equation}
\label{eqn:U_ageostrophic}
\frac{U_\text{ageo}}{L} \sim \frac{W}{H} \quad.
\end{equation}
Since $U_\text{ageo}\sim U_\text{rms}R_0$ where $R_0$ is the Rossby number, $U_\text{ageo}$ can be rewritten as
\begin{equation}
\label{eqn:U_ageostrophic_and_U}
U_\text{ageo} \sim \frac{U^2_\text{rms}}{fL} = \frac{U^2_\text{rms}}{2\Omega\sin\varphi L} \quad.
\end{equation}
where $\varphi$ is the latitude. Combining Equations~(\ref{eqn:U_ageostrophic}) and~(\ref{eqn:U_ageostrophic_and_U}) would give
\begin{equation}
\label{eqn:W}
W = \frac{U^2_\text{rms}H}{2\Omega\sin\varphi L^2}\sim \frac{U^2_\text{rms}H}{\Omega L^2} \quad.
\end{equation}

The settling timescale $\tau_s$ can be calculated in the same format as Equation~(\ref{eqn:tw}),
\begin{equation}
\label{eqn:ts}
\tau_s = \frac{H}{V_s} \quad,
\end{equation}
where $V_s$ is the settling velocity and is given as
\begin{equation}
\label{eqn:Vs}
V_s = \frac{2{\beta}r^2g({\rho}_p - \rho)}{9\eta} \quad,
\end{equation}
where $r$ is the particle radius, $\rho_p$ is the particle density, $\rho$ is the air density, $g$ is the planet's surface gravity and $\eta$ is the viscosity of gas. $\beta$ is the Cunningham slip factor which corrects the Stokes Law in the atmospheric region where the mean free path of air $\lambda_{\mathrm{MFP}}$ is comparable or larger than the particle radius (i.e., Knudsen number $K_\mathrm{N}>>1$). This usually happens in low pressure regimes. This factor takes into account the `slipping' of gas over the particle surface, reducing the drag force predicted by Stokes' Law. Following \citet{Li_and_Wang_2003} and \citet{Spiegel_etal2009}, the Cunningham slip factor $\beta$ can be expressed as
\begin{equation}
\label{eqn:cunning_slip_correction}
\beta = 1 + K_\mathrm{N} \left( 1.256 + 0.4e^{-1.1/K_\mathrm{N}}\right) \quad.
\end{equation}
$K_\mathrm{N}$ is described as 
\begin{equation}
\label{eqn:Kn}
K_\mathrm{N} = \frac{\lambda_{\mathrm{MFP}}}{r}\quad.
\end{equation}
$\lambda_{\mathrm{MFP}}$ is expressed as \citep{Chapman_and_Cowling_1970}
\begin{equation}
\label{eqn:MFP}
\lambda_{\mathrm{MFP}} = \frac{k_\mathrm{B}}{\sqrt{2}{\pi}d^2_{\mathrm{H_2}}} \frac{T}{p} \quad,
\end{equation}
where $T$ and $p$ are the temperature and pressure in the atmosphere, respectively, and $d_{\mathrm{H_2}}$ is the diameter of gas molecule. Following \citet{Parmentier_etal2013} and \citet{Steinrueck_etal2021}, the viscosity $\eta$ of pure molecular hydrogen is assumed. Its analytical formula is given by \citet{Rosner_2000},
\begin{equation}
\label{eqn:MFP}
\eta = \frac{5}{16} \frac{\sqrt{{\pi}mk_BT}}{{\pi}d^2_{\mathrm{H_2}}} \frac{(k_\mathrm{B}T/\epsilon)^{0.16}}{1.22} \quad,
\end{equation}
where $\epsilon$ is the depth of the Lennard-Jones potential well, which equals $59.7k_\mathrm{B}$\,K, and $d_{\mathrm{H_2}}$ is assumed to be $2.827\times10^{-10}$\,m for an \ce{H2}-dominated atmosphere.

As the two limbs experience radiation pressure \citep{Owen_etal2025}, we follow the treatment outlined in \cite{Fu_etal2025}. We first consider the ratio $\beta_\text{rad}$ between the magnitude of the force due to radiation pressure $F_\text{rad}$ and gravity $F_\text{g}$, 
\begin{equation}
\label{eqn:beta_rad}
\beta_\text{rad} = \frac{F_\text{rad}}{F_\text{g}} = \frac{\pi{r^2}Q_\text{pr}(r)F_\ast/c}{4\pi{r^3}\rho{g}/3} = \frac{3Q_\text{pr}(r)F_\ast}{4r{\rho_\text{p}}cg} \quad,
\end{equation}
where $F_\ast$ is the unattenuated stellar flux received by the planet and $c$ is the speed of light. $Q_\text{pr}(r)$ is the radiation pressure efficiency factor depending on the particle size. $Q_\text{pr}(r)$ can be calculated as 
\begin{equation}
\label{eqn:Qpr}
Q_{\text{pr}}(r) = \frac{\int \left[ Q_{\text{ext},\lambda}(r) - \langle g_\lambda(r) \rangle Q_{\text{sca},\lambda}(r) \right]B_\lambda(T_\ast) d\lambda}{\int B_\lambda(T_\ast) d\lambda} \quad,
\end{equation}
where $Q_{\text{ext},\lambda}(r)$ and $Q_{\text{sca},\lambda}(r)$ are the extinction and scattering efficiency at wavelength $\lambda$, $\langle g_\lambda (r)\rangle$ is the asymmetry parameter and $B_\lambda$ is the Planck function at the stellar effective temperature $T_\ast$ \citep{Bohren_and_Huffman_1983}. To include the acceleration of particles due to radiation pressure, $(1+\beta_\text{rad})g$ is used to replace $g$ from Equation~(\ref{eqn:Vs}). We stress that this treatment neglects the effect of non-zero incidence angle of stellar radiation relative to the local gravity vector, as well as the attenuation of shortwave flux at deeper pressures. As a result, the effect of radiation pressure is likely overestimated in our calculations. Further discussions on the effect of such approximation within this framework is presented in Sec.~\ref{sec:discussions}.


Combining the two downward motions $W$ and $V_\text{s}$ (both quantities defined as positive in the downward direction), we can obtain the resulting downward timescale $\tau_v$ given as
\begin{equation}
\label{eqn:tv}
\tau_v = \left( \frac{W + V_\text{s}}{H} \right)^{-1} = \left ( {\tau_w^{-1}} + {\tau_s^{-1}} \right) ^{-1}\quad.
\end{equation}

Here, we define a dimensionless quantity, $\Psi_\text{HALD}$ (Haze Asymmetric Limb Distribution), given as
\begin{equation}
\label{eqn:Rvat}
\Psi_\text{HALD} = \frac{\tau_v}{\tau_a} \quad,
\end{equation}
which characterises whether vertical transport and settling or horizontal advection dominates the motion of haze particles, thereby determining the tendency of a higher or comparable haze concentration over the morning limb compared to that over the evening limb. If $\tau_v = \tau_a$, the particles would just reach the morning limb from the evening limb (Equation~(\ref{eqn:ta})) while traversing vertically over one scale height (Equation~(\ref{eqn:tw}) and~(\ref{eqn:ts})). Where $\Psi_\text{HALD}$$<1$ indicates efficient downward transport (small $\tau_v$) through downward vertical wind and/or gravitational settling. Haze particles would therefore settle efficiently before travelling from the evening limb and reaching the morning limb. Whereas, $\Psi_\text{HALD}$$>1$ indicates efficient horizontal transport (small $\tau_a$) and particles reaching the morning limb before settling. Here, we note that horizontal advection should comprise multiple circulation components alongside the jet, such as divergent and eddy rotational motions \citep{Hammond_and_Lewis_2021}. The atmospheric timescales, features and stellar parameters also vary across hot-Jupiters. However, the assumptions made in this Section only affect the results quantitatively but do not alter the qualitative trends presented by this work. Further discussions on how changing the assumed variables would modify our results quantitatively are presented in Sec.~\ref{sec:discussions}.

\subsubsection{1D Kinematic Model}
\label{subsec:1d_kinematic_model}

To provide a more qualitative interpretation of $\Psi_\text{HALD}$, we develop a simple 1D kinematic model describing the transport of haze particles from the evening to morning limb. The model incorporates the competing effects of horizontal advection and vertical removal, exploring how the resulting limb asymmetry varies as a function of $\Psi_\text{HALD}$. For simplicity, we do not attempt to explicitly represent the trapping of particles within the nightside gyres and other localised circulation features through analytical theories, as their morphology, locations and strength vary among hot-Jupiters. Incorporating such behaviour would require a detailed 3D circulation structure and is therefore not readily captured by a 1D framework. Instead, we focus on the first-order transport between the two planetary limbs.

We first define the 1D (horizontal-only) generic material derivative equation as 
\begin{equation}
\label{eqn:1D_material_derivate}
\frac{D\chi(s)}{Dt} = \frac{\partial\chi(s)}{\partial t} + u\frac{\partial\chi(s)}{\partial s}\quad,
\end{equation}
where $D/Dt$ is the material derivative, $s$ is the distance from the evening limb to any point between the evening and the morning limb, and $u$ is the horizontal wind, which will be expressed in $U_\text{rms}$ for the rest of this work. Assuming that the haze particles are removed through vertical downwelling and settling over the nightside, Equation~(\ref{eqn:1D_material_derivate}) can be expressed as
\begin{equation}
\label{eqn:1D_material_derivate_tau_v}
\frac{D\chi(s)}{Dt} = -\frac{\chi(s)}{\tau_v}\quad.
\end{equation}
Assuming that the system has reached steady state, where $\partial\chi(s)/\partial t=0$, Equations~(\ref{eqn:1D_material_derivate}) and~(\ref{eqn:1D_material_derivate_tau_v}) can be combined to give
\begin{equation}
\label{eqn:1D_material_derivate_steady_state}
U_\text{rms}\frac{\partial\chi(s)}{\partial s} = -\frac{\chi(s)}{\tau_v}\quad.
\end{equation}

Here we define a normalised horizontal coordinate, $x = s/\pi R_p$, where the evening limb corresponds to $x = 0$ and morning limb corresponds to $x = 1$. Since $U_\text{rms}/\pi R_p = 1/\tau_a$, Equation~(\ref{eqn:1D_material_derivate_steady_state}) can therefore be rewritten as
\begin{equation}
\label{eqn:1D_material_derivate_steady_state_hald}
\frac{\partial\chi(x)}{\partial x} = -\frac{\chi(x)}{\Psi_\text{HALD}}\quad.
\end{equation}
Integrating Equation~(\ref{eqn:1D_material_derivate_steady_state_hald}) and defining the boundary condition of $\chi(x = 0) = \chi_e$ where $\chi_e$ is the haze mass mixing ratio over the evening limb, the 1D kinematic model gives the solution 
\begin{equation}
\label{eqn:1D_kinematic_model}
\chi = \chi_e \exp \left( -\frac{x}{\Psi_\text{HALD}} \right)\quad.
\end{equation}
The ratio between the haze mass mixing ratio over the morning limb $\chi_m$ to that over the evening limb $\chi_e$ is therefore 
\begin{equation}
\label{eqn:1D_kinematic_model_ratio}
\frac{\chi_m}{\chi_e} = \exp \left( -\frac{1}{\Psi_\text{HALD}} \right)\quad.
\end{equation}
In other words, Equation~(\ref{eqn:1D_kinematic_model}) shows that prior to considering the trapping effect from the nightside gyres, the values of $\Psi_\text{HALD}$ determines how the haze mass mixing ratio decreases exponentially across the nightside hemisphere. For $\Psi_\text{HALD}<<1$, $\chi_m$ would decrease rapidly as vertical removal dominates. On the other hands for $\Psi_\text{HALD}>>1$, $\chi_m$ would be similar to $\chi_e$ as horizontal mixing dominates.

\subsection{Numerical Model}
\label{subsec:numerical_model}
In this work, we use a 3D GCM, the UK Met Office Unified Model (UM), paired with a haze parameterisation scheme, to simulate the advection, settling and radiative effect of haze particles with varying radii in the atmosphere of WASP-39b. A full description of the haze model, haze properties, and the GCM can be found in \Mtwentyfive\ and a summary presented here. 

\subsubsection{Haze Model}
\label{subsubsec:haze_model}
The haze model builds upon the work from \citet{Steinrueck_etal2021,Steinrueck_etal2023}, describing a haze production profile as a log-normal distribution in pressure, and a haze removal profile through a boundary condition. The model setup in this work is nearly equivalent to that in \Mtwentyfive\ where the haze particles are treated as radiatively active species (unless stated otherwise), except for the column-integrated haze mass production rate, $F_{0}$, and the median of distribution (pressure level in which haze production peaks), $p_m$. For $F_{0}$ we vary between 3$\times$10$^{-15}$, 1$\times$10$^{-14}$, and 1$\times$10$^{-13}$\,\massflux. For the lower bound on $F_{0}$, we follow \citet{Arfaux_and_Lavvas_2024}, who use 3$\times$10$^{-15}$\,\massflux in their 1D study coupling cloud formation in the atmosphere of WASP-39b, with photochemical haze acting as a cloud condensing species. An earlier study by \citet{Arfaux_and_Lavvas_2022} have also suggested $F_{0}$ to be between 1$\times$10$^{-16}$--1$\times$10$^{-15}$\,\massflux in WASP-39b. For the upper bound, we have adopted the value of 1$\times$10$^{-13}$\,\massflux as previous 3D models have already studied a much higher production rate (1$\times$10$^{-12}$\,\massflux in \Mtwentyfive\ and 2.5$\times$10$^{-12}$\,\massflux in \citealt{Espinoza_2024} and \citealt{Steinrueck_etal2025}). We therefore select parameters that fill the interval spanned by these studies and explore the sensitivity of limb haze features to varying haze mass production rates. For $p_m$ we opted for a value of 0.001\,mbar, motivated by the results of \citet{Arfaux_and_Lavvas_2024}, who show that the haze production peaks at $\sim$0.001\,mbar in both terminators of WASP-39b (see their Fig.\,3) using a self-consistent 1D radiative convective model that also takes into account disequilibrium chemistry and haze microphysics. We note that this pressure is well below the typical $\sim$1\,mbar pressures probed by transmission spectroscopy. 

\begin{table}
\centering
	\caption{Parameter space of simulations.}
	\label{tab:haze_model_parameter}
        \begin{tabular}{ |l|l| }
        \hline
        Parameter & Value \\
        \hline
        $F_{0}$ [\massflux] & 3$\times$10$^{-15}$, 1$\times$10$^{-14}$, 1$\times$10$^{-13}$ \\
        \hline
        Radii [nm] & 1.5, 5, 15, 25, 50 \\
		\hline
	\end{tabular}
\end{table}

Following \Mtwentyfive, the haze particles are assumed to be spherical. Mie theory is used by the radiative transfer code in the 3D GCM (see Sec.~\ref{subsubsec:climate_model}) to calculate the absorption, scattering coefficients and asymmetry parameters of the haze particles \citep{Bohren_etal2008}. The particle mean radius varies from 1.5, 5, 15, 25, to 50\,nm, motivated by the results from \citet{Arfaux_and_Lavvas_2024} who show that the mean radius of both terminators varies between 1\,nm and 100\,nm for pressures $<10^3$\,mbar. The parameters explored in our simulations are summarised in Tab.~\ref{tab:haze_model_parameter}. The haze particles are assumed to be soot-like in composition as they tend to form under high temperature and have been commonly used by the exoplanet community, with the optical properties taken from \citet{Lavvas_and_Koskinen_2017} (see Fig.\,1(i-ii) in \Mtwentyfive). 

\subsubsection{Climate Model}
\label{subsubsec:climate_model}

The UM uses the the dynamical core, Even Newer Dynamics for General atmospheric modelling of the environment (ENDGame), which adopts a semi-implicit semi-Lagrangian scheme to solve the non-hydrostatic, full deep-atmosphere equations of motion with varying gravity within the atmosphere \citep[see][for discussion]{Wood_etal2014,Mayne_etal2014a,Mayne_etal2014b,Mayne_etal2017,Mayne_etal2019}. The UM uses the 2-stream radiative transfer scheme, the ``Suite of Community RAdiative Transfer codes based on \citet{Edwards_and_Slingo_1996}'' (Socrates), to calculate the absorption and scattering coefficients and asymmetry parameter of the haze particles. It also solves for gaseous absorption from \ce{H2O}, \ce{CH4}, \ce{CO}, \ce{Cs}, \ce{K}, \ce{Li}, \ce{Na}, \ce{NH3}, \ce{Rb} and collision-induced absorption from \ce{H2}-\ce{H2} and \ce{H2}-\ce{He} from the ExoMol line lists \citep{Tennyson_etal2016} using the correlated-\textit{k} method \citep{Amundsen_etal2014,Amundsen_etal2016,Amundsen_etal2017,Goyal_etal2020}.

The stellar and planetary parameters of WASP-39b are the same as \Mtwentyfive\ (see their Tabs.\,1 and 2), except for the solar metallicity in which \Mtwentyfive\ adopted the value of 1$\times$solar and here we have used 10$\times$solar, following the hazy and the cloudless equilibrium chemistry and transport induced disequilibrium chemistry simulations presented in \citet{Espinoza_2024}. The adopted metallicity is used to derive the bulk thermodynamic properties of the atmosphere, including the specific gas constant and specific heat capacity, which are then passed to the dynamical and radiative components of the GCM. This results in values of 3165\,J\,K$^{-1}$\,kg$^{-1}$ and 1.15$\times$10$^{4}$\,J\,K$^{-1}$\,kg$^{-1}$, respectively. The simulations do not include the effect of clouds, kinetic chemistry, and UV photolysis but use the analytical scheme from \citet{Burrows_and_Sharp_1999} to calculate the alkali metal and chemical equilibrium abundances of \ce{CH4}, \ce{CO}, \ce{H2O} and \ce{NH3} \citep[see][for more details]{Amundsen_etal2016}. We note that the parameterisation scheme from \citet{Amundsen_etal2016} assumes approximately a solar elemental abundance, and might not fully reflect the adopted 10$\times$solar metallicity. However, another motivation for adopting the 10$\times$solar metallicity configuration is that it also allows for the model top of our simulation to be extended to pressure of $\leq10^{-3}$\,mbar while maintaining numerical stability. By comparison, the 1$\times$solar simulations presented in \Mtwentyfive\ were limited to a model top of $\sim$0.05\,mbar under stable conditions. This is particularly important for the present study because gravitational settling is strongest at the lowest pressures. Extending the model domain therefore allows us to better probe the upper atmosphere, where the transition between higher haze concentration over the morning and the evening limb might occur. Furthermore, since this work focuses on exploring how particle size affects the resulting asymmetrical limb distribution of haze, we defer a more detailed and accurate chemical network coupled to photochemical haze for this planet to future work. All simulations are run for 2000\,Earth days such that a quasi-equilibrium is reached in the upper atmosphere at pressures $\lesssim10^3$\,mbar (region bounded by haze production and removal, see Sec.~\ref{subsubsec:haze_model}), diagnosed through analysing the percentage change of total haze mass mixing ratio compared to initial total haze mass mixing ratio, and the global top-of-atmosphere net radiative flux fluctuating less than 1\%. Finally, the last 100\,Earth days are temporally averaged for all data analysis. 


\section{Results}
\label{sec:results}
We present our results in two parts, starting with Sec.~\ref{subsec:advection_vs_settling} where we introduce the HALD framework that balances advection and settling, enabling comparisons with GCM simulations and past work that omit radiation pressure. Afterwards we introduce Sec.~\ref{subsec:advection_vs_settling_vs_radiation_pressure} and extend the framework to include the effects of radiation pressure.

\subsection{Advection vs Settling}
\label{subsec:advection_vs_settling}

\begin{figure*}
\centering
 \includegraphics[width=140mm]{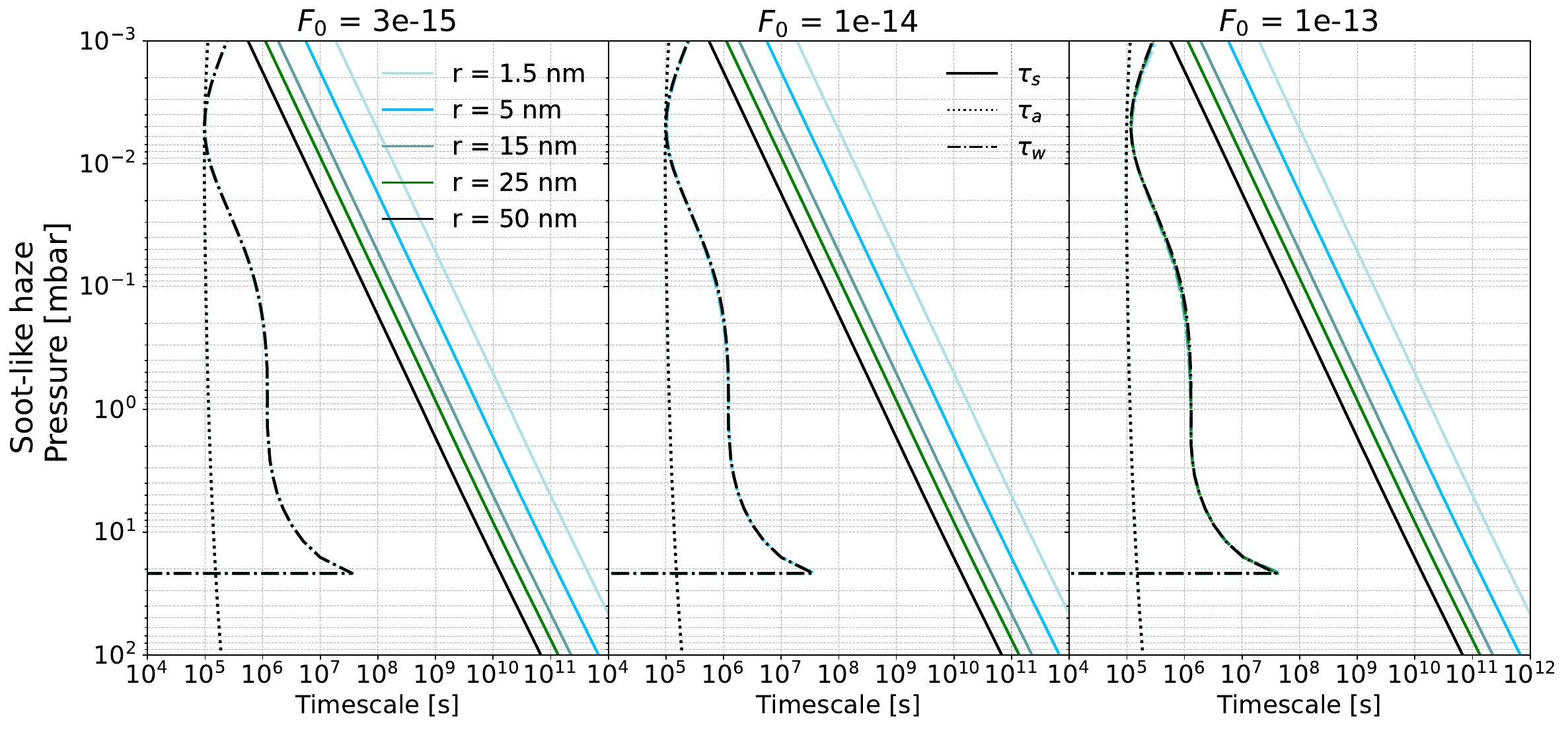}
 \caption{Gravitational settling ($\tau_s$ -- solid), horizontal advection ($\tau_a$ -- dotted), and vertical advection ($\tau_w$ -- dash-dotted) timescales of soot-like haze across different haze mass production rates and radii, calculated from UM simulations.}
 \label{fig:tw_ts_ta_MAD}
\end{figure*}

\begin{figure}
\centering
 \includegraphics[width=85mm]{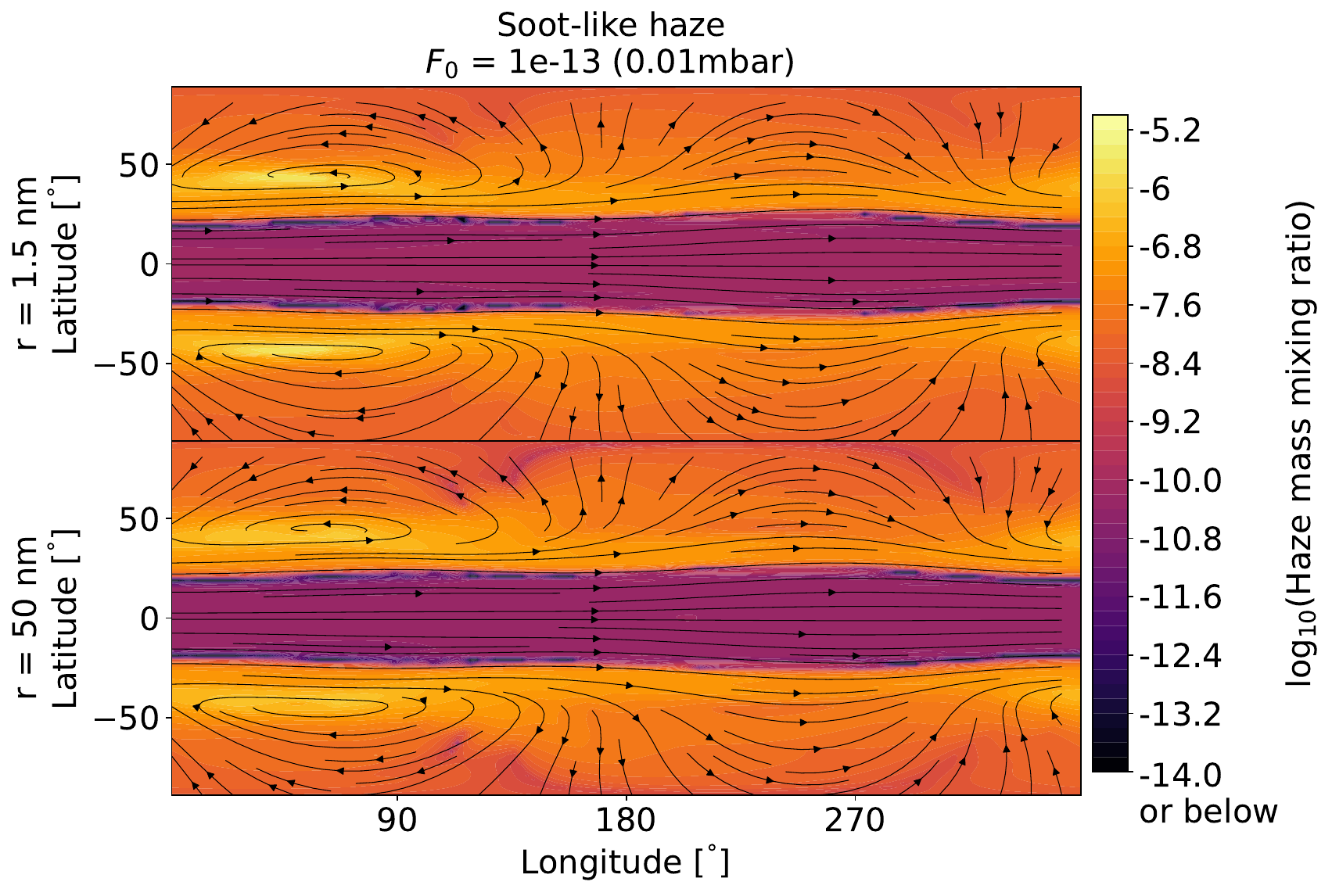}
 \caption{Spatial distribution of haze mass mixing ratio (contour) and horizontal flow (streamlines) in WASP-39b, with $F_0$$=$ 1$\times$10$^{-13}$\,\massflux at pressure 0.01\,mbar with particle radii of 1.5\,nm (top) and 50\,nm (bottom).}
 \label{fig:f1e-13_MMR_1e-5bar_MAD}
\end{figure}

\begin{figure}
\centering
 \includegraphics[width=90mm]{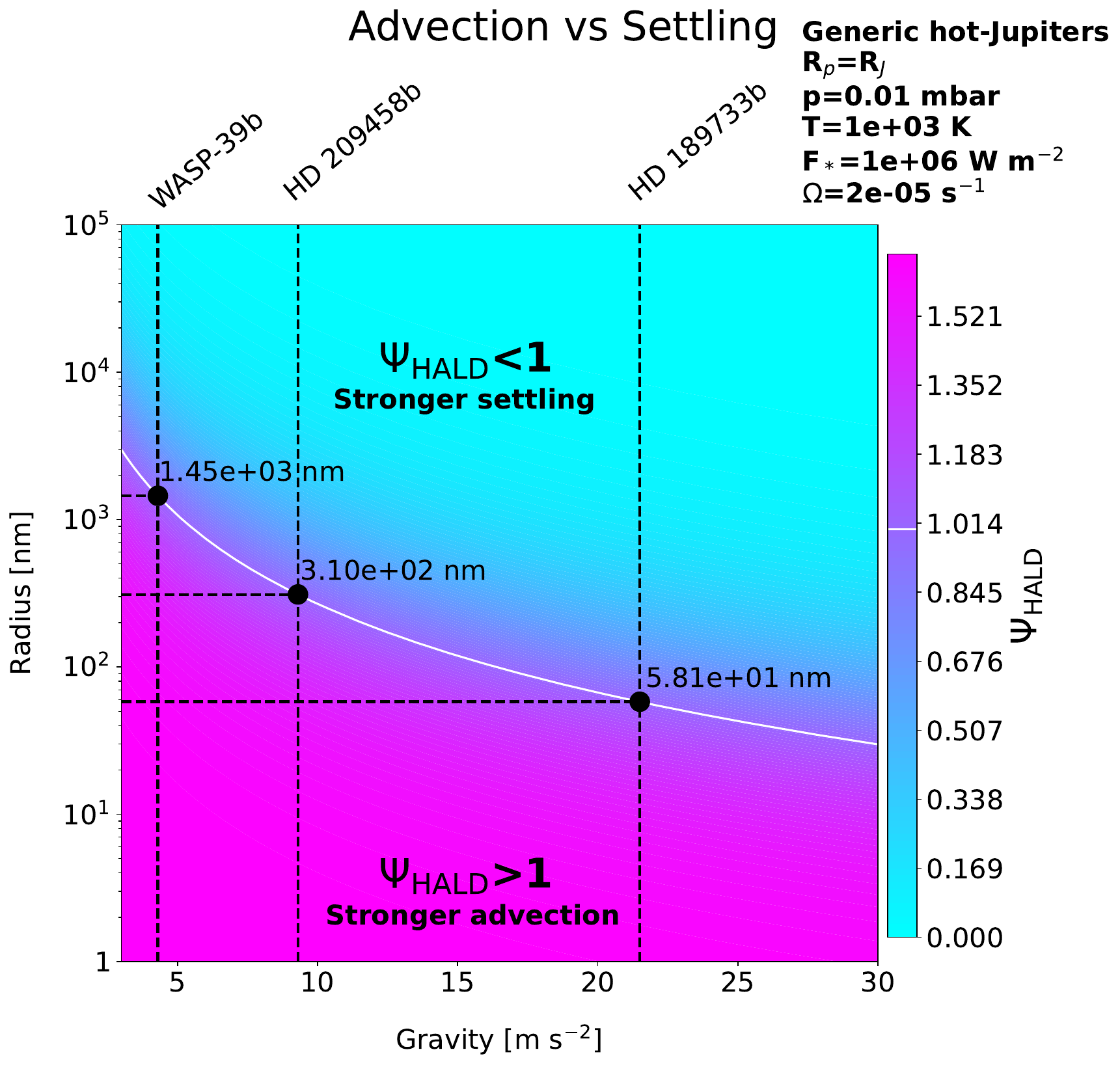}
 \caption{Values of $\Psi_\text{HALD}$ ($\tau_v/\tau_a$) = 1 as a function of particle radii and planetary surface gravities, assuming generic atmosphere parameters of a standard hot-Jupiter with radius of $1\times R_J$ where R$_J$ is Jupiter radius at a pressure $=$ 0.01\,mbar, temperature $=$ 1000\,K, $F_\ast=1\times10^6$\,W\,m$^{-2}$ and rotation rate $=$ 2$\times$10$^{-5}$\,s$^{-1}$. The solid line indicates the contour where $\Psi_\text{HALD}=1$. The surface gravities of three well-studied hot-Jupiters, including HD\,189733b, HD\,209458b and WASP-39b are marked, along with their corresponding maximum particle radius for which haze particles can survive the hemispheric transport and reach the morning limb before vertical removal dominates are indicated. Note that we expect that particles reaching the morning limb may subsequently accumulate within the nightside gyres, thereby resulting in a higher or comparable concentration over the morning limb relative to the evening limb.}
 \label{fig:metric_limb_prediction_generic_1e-5mbar_1000K}
\end{figure}

Fig.~\ref{fig:tw_ts_ta_MAD} shows the calculated $\tau_a$ (Equation~(\ref{eqn:ta})), $\tau_w$ (Equation~(\ref{eqn:tw})) and $\tau_s$ (Equation~(\ref{eqn:ts})) diagnosed from our 3D simulations with varying haze mass production rates and radii in WASP-39b. Fig.~\ref{fig:tw_ts_ta_MAD} shows that $\tau_s$ differs significantly between each radius, compared to the almost unchanging $\tau_a$ and $\tau_w$. The larger the particle, the smaller the $\tau_s$. $\tau_s$ is also larger at higher pressures, indicating weak gravitational settling deeper in the atmosphere. 

Fig.~\ref{fig:f1e-13_MMR_1e-5bar_MAD} shows the comparison of the spatial distribution of haze, with the horizontal flow over-plotted, at a pressure of 0.01\,mbar with $F_0 = 1\times10^{-13}$\,\massflux between the case with radius of 1.5\,nm and 50\,nm (the two ends of the parameter space). The simulations with other values of $F_0$ show similar distributions and differ only through presentation of a reduced haze concentration. We therefore present only the most illustrative cases. Fig.~\ref{fig:f1e-13_MMR_1e-5bar_MAD} shows a lower haze concentration in both cases compared to \Mtwentyfive\ where they opted for $F_0 = 1\times10^{-12}$\,\massflux. This leads to a weaker radiative forcing in the atmosphere compared to \Mtwentyfive, resulting in nearly equivalent horizontal flow (Fig.~\ref{fig:f1e-13_MMR_1e-5bar_MAD}), with $\tau_a$ and $\tau_w$ being almost identical among all cases in Fig.~\ref{fig:tw_ts_ta_MAD}. Fig.~\ref{fig:f1e-13_MMR_1e-5bar_MAD} also shows that the simulation with larger haze particles in general presents a slightly lower haze concentration, due to the increased gravitational settling as shown in Fig.~\ref{fig:tw_ts_ta_MAD}. Yet, the pattern of haze distribution among all cases remains similar, with more haze concentrated over the morning limb ($\sim90^\circ$ longitude) despite the almost two orders of magnitude differences in gravitational settling timescale. Furthermore, from Fig.~\ref{fig:tw_ts_ta_MAD}, $\tau_a$ is $\sim$1--6 orders of magnitude smaller than $\tau_s$ across pressure levels, meaning advection is dominating over gravitational settling throughout almost the entire atmosphere. This is due to the low surface gravity of WASP-39b (4.3\,\g -- see Tab.\,2 from \Mtwentyfive) which weakens the gravitational settling significantly. This leads to an efficient transport of haze to the morning limb regardless of the particle sizes in our simulations.

To explore the impact of surface gravities, we calculate the $\Psi_\text{HALD}$ for a range of haze particle radii (1--10$^5$\,nm) and surface gravities (3--30\,\g), assuming a generic hot-Jupiter with planetary radius of $1\times R_J$ where R$_J$ is Jupiter radius, temperature $=$ 1000\,K, $F_\ast=1\times10^6$\,W\,m$^{-2}$, rotation rate $=$ 2$\times$10$^{-5}$\,s$^{-1}$, at pressure $=$ 0.01\,mbar. The contour where $\Psi_\text{HALD}=1$ is shown in Fig.~\ref{fig:metric_limb_prediction_generic_1e-5mbar_1000K}. It corresponds to the estimated maximum radius for which haze particles can survive the hemispheric transport and reach the morning limb before vertical removal dominates. Throughout this work we assume that particles reaching the morning limb may subsequently accumulate within the nightside gyres (see Figs. 6, 9 and 12 in \Mtwentyfive), a process not explicitly represented by the analytical framework, thereby resulting in a higher or comparable concentration over the morning limb relative to the evening limb (see Sec.~\ref{subsec:analytical_theory}). The surface gravities of three well-studied hot-Jupiters which have been previously modelled by 3D GCMs are also marked on Fig.~\ref{fig:metric_limb_prediction_generic_1e-5mbar_1000K}, assuming the presence of haze. They include HD\,189733b \citep[$g=21.5$\,\g,][]{Steinrueck_etal2021,Steinrueck_etal2023}, HD\,209458b ($g=9.3$\,\g, \Mtwentyfive), and WASP-39b ($g=4.3$\,\g, this work and \Mtwentyfive). 

For HD\,189733b, Fig.~\ref{fig:metric_limb_prediction_generic_1e-5mbar_1000K} estimates the maximum particle radius $r_\text{max}(p = 0.01\,\text{mbar})$, for which haze particles can survive transport to the morning limb, subsequently be trapped within the gyres and result in a higher or comparable haze concentration over the morning limb, to be 58.1\,nm. Compared with \citet{Steinrueck_etal2021}, who vary the radius between 1\,nm, 3\,nm, 10\,nm, 30\,nm, 100\,nm, 300\,nm, and 1000\,nm, at the same pressure of 0.01\,mbar, their simulation results show that the maximum radius lies <100\,nm in order to see more haze over the morning limb at pressure $=$ 0.01\,mbar (see their Fig.\,2). For larger radii at the same pressure level, \citet{Steinrueck_etal2021} show that either both limbs have almost the same concentration (when $r=$ 100\,nm) or the evening limb has a higher concentration (when $r=$ 300\,nm and 1000\,nm). We note, however, that their 3D simulations capture the westward flow of haze particles from the day side. If the effect of this westward flow were neglected, the maximum particle size allowing comparable or higher haze concentration on the morning limb would reduce in their simulations (see Sec.~\ref{sec:discussions} for further discussions). 


Contrary to HD\,189733b, the lower surface gravities of both HD\,209458b and WASP-39b (9.3\g and 4.3\g, respectively) result in weaker gravitational settling. Consequently, efficient horizontal transport and the higher concentration of haze over the morning limb can still occur despite a large particle size. This can be seen in Fig.~\ref{fig:metric_limb_prediction_generic_1e-5mbar_1000K} which estimates the maximum radius $r_\text{max}(p = 0.01\,\text{mbar})$ to be 310\,nm and 1450\,nm for HD\,209458b and WASP-39b. We compare our results with \Mtwentyfive, who perform simulations with the UM with the haze radius of 1.5\,nm in HD\,209458b. \Mtwentyfive\ show that the haze particles are trapped efficiently within the nightside gyres, resulting in more haze over the morning limb across all pressure levels. We also compare our work with \citet{Steinrueck_etal2025}, who perform SPARC/MITgcm simulations with a particle radius of 30\,nm for WASP-39b and also showed a higher build-up of haze over the morning limb at a pressure of 0.01\,mbar. Our work here, which explores particle radii of up to 50\,nm, shows similar results (see Fig.~\ref{fig:f1e-13_MMR_1e-5bar_MAD}). 

\begin{figure}
\centering
 \includegraphics[width=80mm]{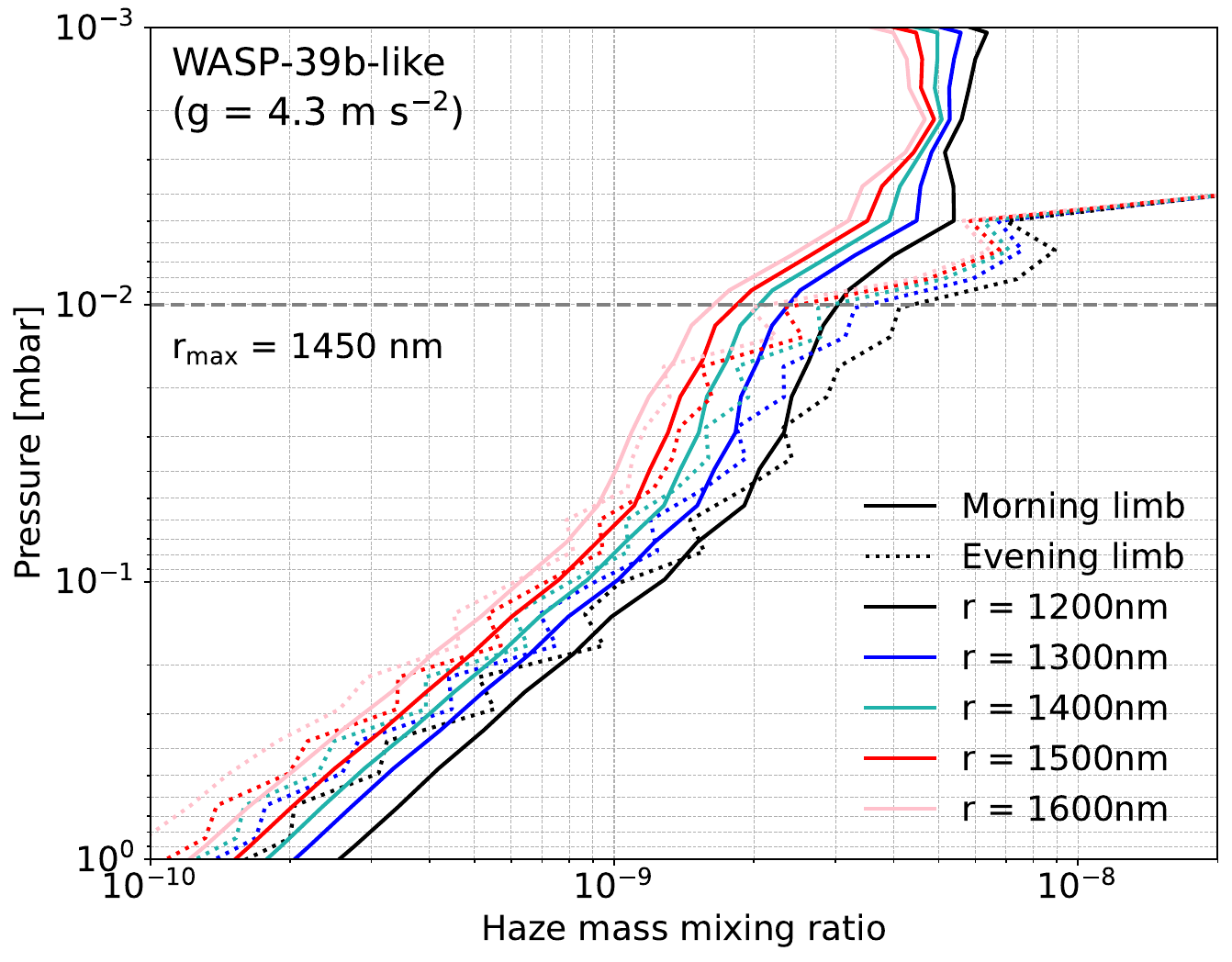}
 \caption{Haze mass mixing ratio between the morning (solid) and evening limb (dotted) for a WASP-39b-like atmosphere at g = 4.3\g, with $F_0$$=$ 1$\times$10$^{-13}$\,\massflux and particle radii of 1200, 1300, 1400 1500 and 1600\,nm. The dashed horizontal line indicates the pressure level $=$ 0.01\,mbar at which the framework approximates the maximum radius $r_\text{max}(p = 0.01\,\text{mbar})$ to be 1450\,nm in order to survive transport to the morning limb. Under the expectation of subsequent trapping within the nightside gyres, this corresponds to the maximum particle radius capable of producing a higher or comparable haze mass mixing ratio over the morning limb than the evening limb at this pressure level. All haze particles in these simulations are treated as radiative inactive passive tracers.}
 \label{fig:g_4_varying_r}
\end{figure}

\begin{figure}
\centering
 \includegraphics[width=80mm]{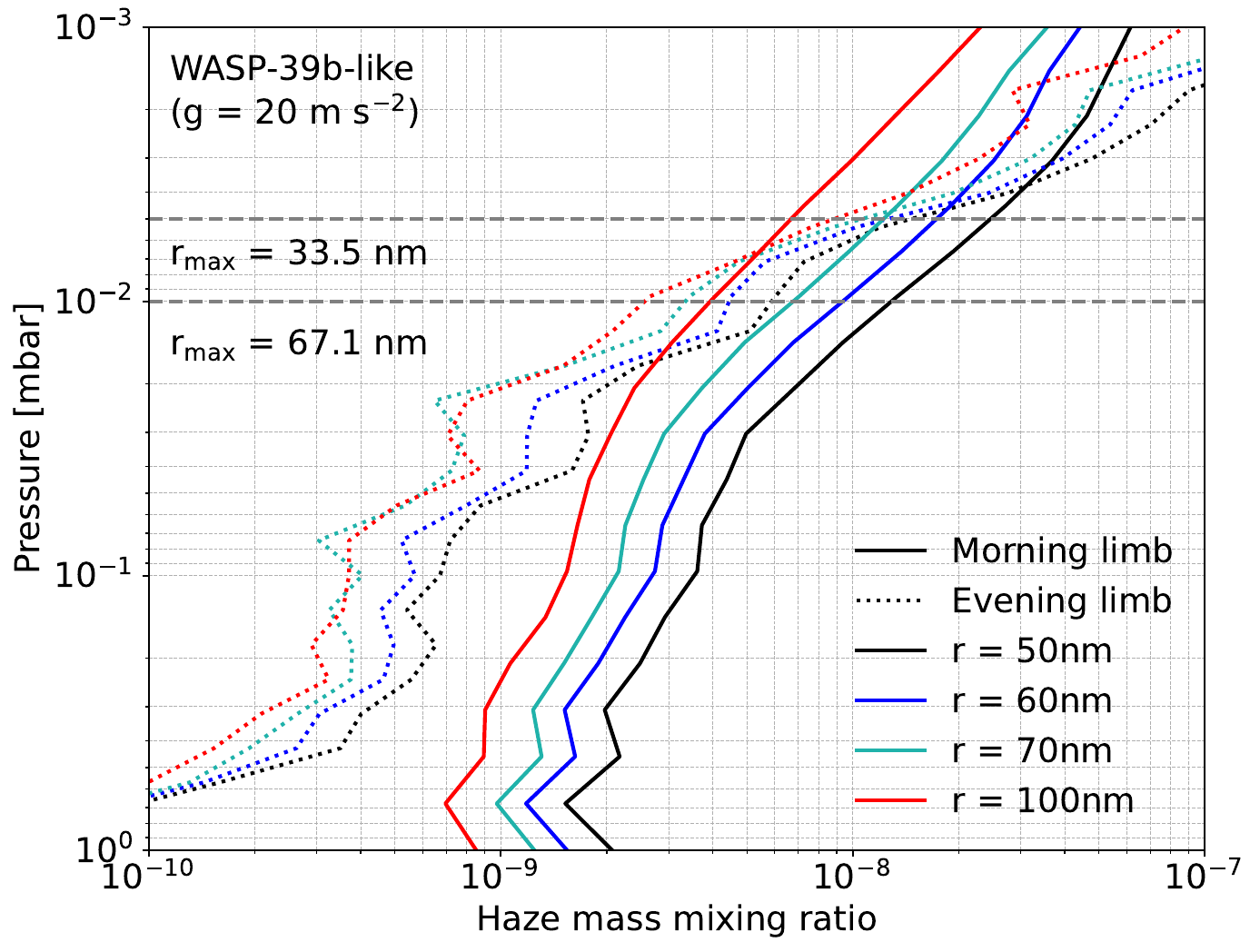}
 \caption{Same format as shown in Fig.~\ref{fig:g_4_varying_r} but at g = 20\g), with $F_0$$=$ 1$\times$10$^{-13}$\,\massflux and particle radii of 50, 60, 70 and 100\,nm. The dashed horizontal line indicates the pressure level $=$ 0.005 and 0.01\,mbar at which the framework approximates the maximum radius to be 33.1 and 67.1\,nm, respectively, in order to survive transport to the morning limb and maintain a higher or comparable haze mass mixing ratio over the morning limb relative to the evening limb at such pressure levels.}
 \label{fig:g_20_varying_r}
\end{figure}

To better test the transition between a higher haze mass mixing ratio over the morning and the evening limb approximated by our framework,  a series of additional simulations have been performed with $F_0$$=$ 1$\times$10$^{-13}$\,\massflux, but through treating the haze particles as radiatively inactive passive tracers, meaning that they have no radiative effect on the planetary atmosphere. These simulations adopt the same  WASP-39b-like atmospheric configuration described in Sec.~\ref{subsec:numerical_model}. However, two surface gravities are considered, including $g=$ 4.3\g\ which corresponds to WASP-39b, and $g=$ 20\g\ which corresponds to a strong-settling regime. For the case with $g=$ 4.3\g, our framework from Fig.~\ref{fig:metric_limb_prediction_generic_1e-5mbar_1000K} approximated the maximum particle radius $r_\text{max}(p = 0.01\,\text{mbar})$ to be 1450\,nm at pressure $=$ 0.01\,mbar, motivating the particle radius within the additional simulations to be set at 1200, 1300, 1400, 1500 and 1600\,nm. For the case with $g=$ 20\g, our framework from Fig.~\ref{fig:metric_limb_prediction_generic_1e-5mbar_1000K} approximated $r_\text{max}(p = 0.01\,\text{mbar})$ to be 67.1\,nm at pressure $=$ 0.01\,mbar, motivating the particle radius within the simulations to be set at be 50, 60, 70 and 100\,nm.

Fig.~\ref{fig:g_4_varying_r} shows the haze mass mixing ratio distribution between the morning and evening limbs from the simulation with $g=$ 4.3\g. At pressure of 0.01\,mbar, Fig.~\ref{fig:metric_limb_prediction_generic_1e-5mbar_1000K} shows that the approximated $r_\text{max}(p = 0.01\,\text{mbar})$ is 1450\,nm for haze particles to survive transport to the morning limb. Assuming subsequent accumulation within the nightside gyres, which is not explicitly modelled in the analytical framework (see Sec.~\ref{subsec:analytical_theory}), these particles can then contribute to a higher or comparable haze mass mixing ratio over the morning limb relative to the evening limb at this pressure level. Fig.~\ref{fig:g_4_varying_r} shows that all simulations show a higher haze mass mixing ratio over the evening limb despite the larger particle sizes. When combined with the simulation results listed in Sec.~\ref{subsec:numerical_model} (radii of 1.5, 5, 15, 25, and 50\,nm), the maximum particle radius above which haze can no longer survive across the hemispheric transport and the morning limb can no longer show a higher mass mixing ratio relative to the evening limb lies within the range $50\ \text{nm}<r_\text{max}(p = 0.01\,\text{mbar})<1200$\,nm. Although the exact transition is not captured amongst the additional simulation sets, Fig.~\ref{fig:g_4_varying_r} shows that the evening limb is only showing a slightly higher haze mass mixing ratio than the morning limb for particle radii $\geq$1200\,nm. This suggests that the GCM-diagnosed maximum radius might lie around the 1000\,nm regime, and is broadly consistent with our framework to the same order-of-magnitude.

Fig.~\ref{fig:g_20_varying_r} shows the haze mass mixing ratio distribution between the morning and evening limbs from the simulation with $g=$ 20\g. At pressure of 0.01\,mbar, Fig.~\ref{fig:metric_limb_prediction_generic_1e-5mbar_1000K} shows that the approximated $r_\text{max}(p = 0.01\,\text{mbar})$ is 67.1\,nm for particles to survive the whole nightside transport and therefore for the morning limb to maintain a higher haze mass mixing ratio. At this pressure level, all simulations, including when $r=100$nm, exhibit a higher haze mass mixing ratio over the morning limb, indicating that 100\,nm$<r_\text{max}(p = 0.01\,\text{mbar})$. Comparing the limbs at a pressure of 0.005\,mbar where gravitational settling is stronger, our framework estimates the $r_\text{max}(p = 0.005\,\text{mbar})$ to be 33.1\,nm. Fig.~\ref{fig:g_20_varying_r} shows that from the 3D climate simulation in the case with $r=100$\,nm, the evening limb shows a higher haze concentration than the morning limb, while in the case with $r=70$\,nm, both limbs show almost identical haze concentrations. In the case when $r=50$ and 60\,nm, the settling becomes weak enough that the morning limb is able to have a higher haze concentration over the evening limb. The 3D simulations therefore place the constraints the maximum radius lies within the range of $60\ \text{nm}<r_\text{max}(p = 0.005\,\text{mbar})<70$\,nm, compared to the analytical approximation of 33.1\,nm. A summary of the comparison between the maximum particle radius predicted by the HALD framework and that diagnosed from the GCM simulations is presented in Tab.~\ref{tab:MAD_comparison}. The above comparison demonstrates that our analytical framework successfully captures the first-order balance between horizontal mixing and vertical removal. Despite the simplifying assumptions embedded within the analytical theories, and under the assumption that haze particles reaching the morning limb subsequently accumulate within the nightside gyres, the framework approximates particle sizes that are broadly consistent with those diagnosed from the GCM simulations.



\begin{figure}
\centering
 \includegraphics[width=85mm]{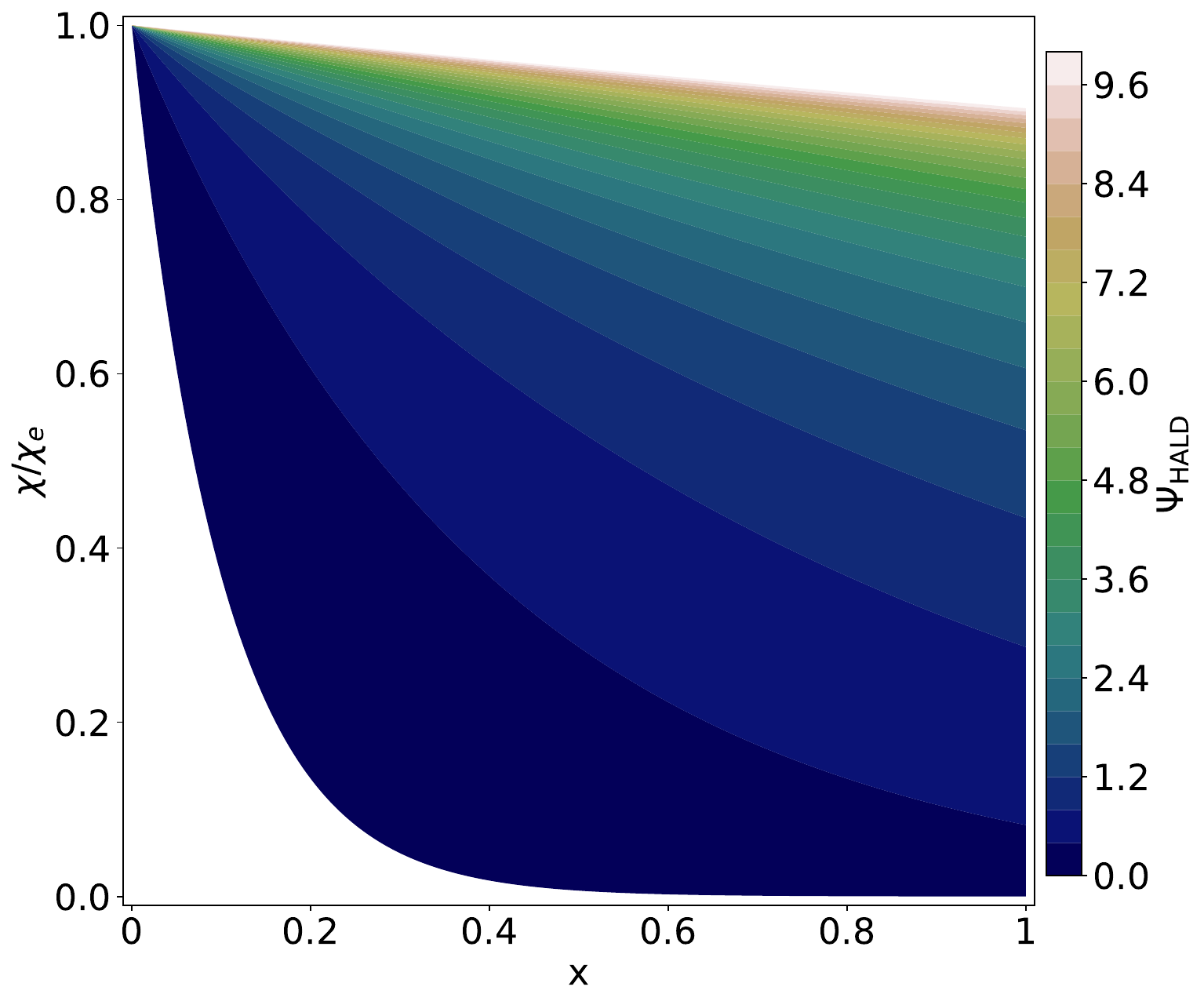}
 \caption{1D kinematic model based on Equation~(\ref{fig:kinematic_model}) which shows the normalised haze mass mixing ratio with respect to the evening limb ($\chi/\chi_e$) between the evening limb ($x=0$) and morning limb ($x=1$) for different values of $\Psi_\text{HALD}$. The model neglects the trapping of haze particles within the nightside gyres and illustrates the first-order impact of the balance between the horizontal transport and vertical removal on the resulting limb asymmetry.}
 \label{fig:kinematic_model}
\end{figure}

To further understand the balance between horizontal mixing and vertical removal captured by the HALD framework, Equation~(\ref{eqn:1D_kinematic_model}) is plotted and shown in Fig.~\ref{fig:kinematic_model} which illustrates the normalised haze mass mixing ratio with respect to the evening limb ($\chi/\chi_e$) between the evening limb ($x=0$) and morning limb ($x=1$) for different values of $\Psi_\text{HALD}$. The model neglects the trapping of haze particles within the nightside gyres (see Sec~\ref{subsec:1d_kinematic_model}), which would result in a build-up of particles and therefore a higher haze mass mixing ratio over the morning limb, but illustrates the first-order impact of the balance between the horizontal transport and vertical removal on the resulting limb asymmetry. As expected, Fig.~\ref{fig:kinematic_model} shows that haze mass mixing ratio decreases exponentially with increasing distance from the evening limb. The depletion of haze particles reduces for increasing $\Psi_\text{HALD}$, reflecting the increasingly efficient horizontal transport associated with $\Psi_\text{HALD}>>1$. On the other hand, $\Psi_\text{HALD}<<1$ indicates strong settling and efficient removal of particles to the deep atmosphere during transport across the whole nightside. Since trapping within the nightside gyres is neglected, this model cannot produce the $\chi/\chi_e>1$. However in this context, the HALD framework and our 1D kinematic model provide a useful starting point for interpreting haze behaviour: when $\Psi_\text{HALD}>1$, corresponding to an efficient horizontal mixing, full 3D GCMs are required to assess the extend of limb asymmetries of haze distribution, whereas when $\Psi_\text{HALD}<1$, corresponding to an efficient settling, haze particles will be strongly biased towards concentrating over the evening limb.


\begin{table*}
\centering
	\caption{Comparison of the estimated maximum radius [nm] predicted by the HALD framework for haze particles to survive transport to the morning limb, and the corresponding transition particle sizes diagnosed from GCM simulations in this work and the literature. Expecting subsequent trapping within the nightside gyres, values predicted by he HALD framework correspond to the maximum particle sizes capable of producing a higher or comparable haze concentration over the morning limb than the evening limb. Note that only particles with radius 1.5\,nm and 30\,nm are presented in \citet{Mak_etal2025} for HD\,209458b and \citet{Steinrueck_etal2025} for WASP-39b, respectively.}
	\label{tab:MAD_comparison}
        \begin{tabular}{|p{4cm}|p{3cm}|p{7cm}|}
        \hline
        Planet & HALD estimation [nm] & GCM simulations [nm] \\
        \hline
        HD\,189733b & 58.1 ($p=0.01$\,mbar) & $30<r_\text{max}(p = 0.01\,\text{mbar})<100$ \citep{Steinrueck_etal2021}\\
        \hline
        HD\,209458b & 310 ($p=0.01$\,mbar)& $r_\text{max}(p = 0.01\,\text{mbar})\geq1.5$ \citep{Mak_etal2025}\\
        \hline
        WASP-39b & 1450 ($p=0.01$\,mbar)& $50<r_\text{max}(p = 0.01\,\text{mbar})<1200$ (This work) \\
                 &      & $r_\text{max}(p = 0.01\,\text{mbar})>30$ \citep{Steinrueck_etal2025} \\
        \\
        WASP-39b-like (g=20\g) & 67.1 ($p=0.01$\,mbar)& $100<r_\text{max}(p = 0.01\,\text{mbar})$ (This work) \\
                               & 33.1 ($p=0.005$\,mbar)& $60<r_\text{max}(p = 0.005\,\text{mbar})<70$ (This work) \\
		\hline
	\end{tabular}
\end{table*}

\begin{figure}
\centering
 \includegraphics[width=90mm]{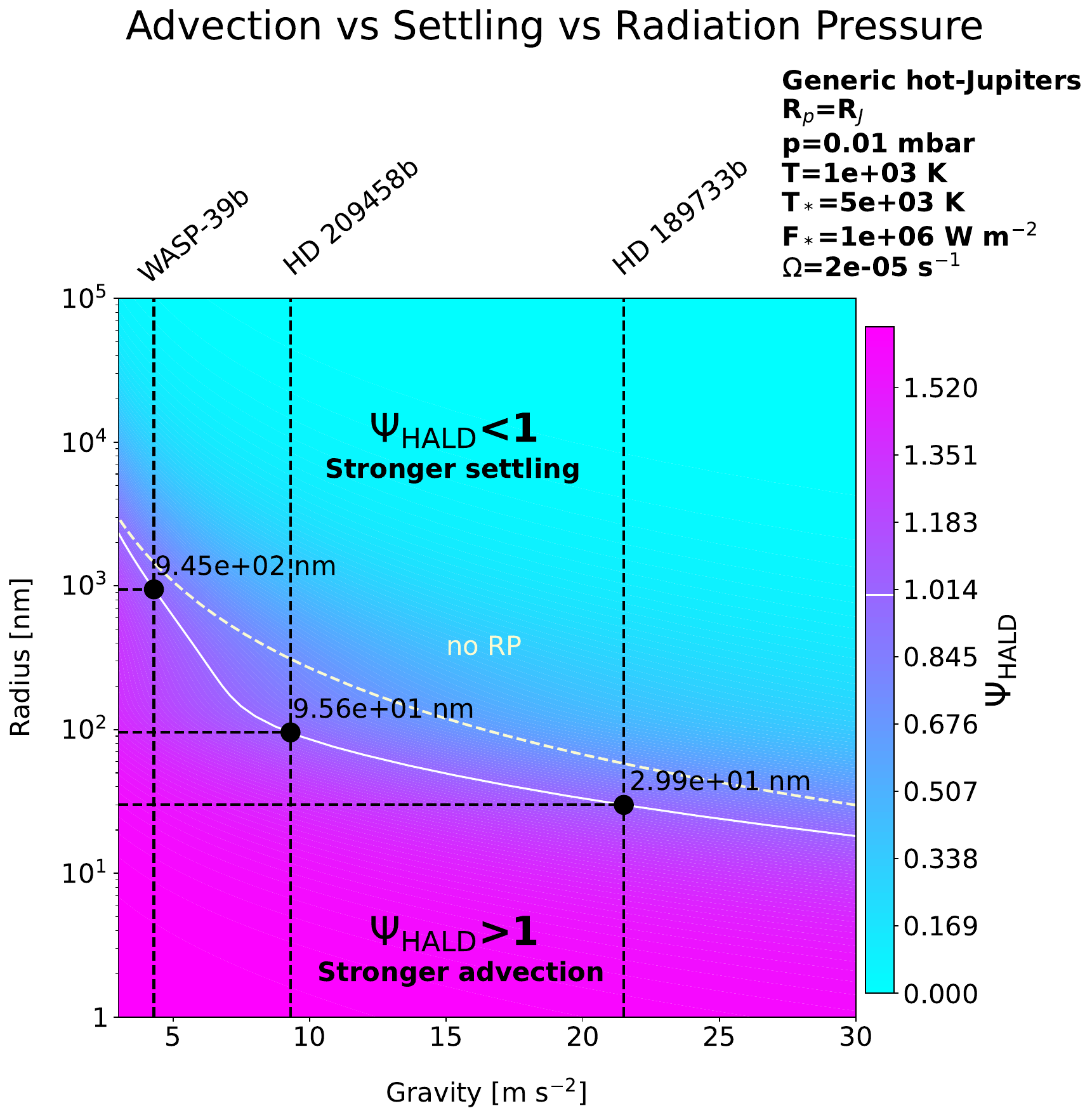}
 \caption{Same format as shown in Fig.~\ref{fig:metric_limb_prediction_generic_1e-5mbar_1000K} but including the effect of radiation pressure assuming $T_\ast=5000$\,K and $F_\ast=1\times10^6$\,W\,m$^{-2}$. The solid line indicates the contour where $\Psi_\text{HALD}=1$ when radiation pressure is included, while the dashed light yellow line shows the corresponding $\Psi_\text{HALD}=1$ boundary without radiation pressure (noted as ``no RP'' on the figure) from Fig~\ref{fig:metric_limb_prediction_generic_1e-5mbar_1000K}.}
 \label{fig:metric_limb_prediction_generic_1e-5mbar_1000K_radiation_pressure}
\end{figure}

\subsection{Advection vs Settling vs Radiation Pressure}
\label{subsec:advection_vs_settling_vs_radiation_pressure}
In this Section, we include radiation pressure in our HALD calculations assuming a Sun-like star with a stellar temperature $T_\ast$ of 5000\,K and stellar flux $F_\ast$ of 1$\times10^6$\,W\,m$^{-2}$. We compute $\Psi_\text{HALD}$ and the contour of $\Psi_\text{HALD}=1$ is presented in Fig.~\ref{fig:metric_limb_prediction_generic_1e-5mbar_1000K_radiation_pressure}, with the same planetary and atmospheric parameters as Fig.~\ref{fig:metric_limb_prediction_generic_1e-5mbar_1000K}. For comparison, we have also included the corresponding $\Psi_\text{HALD}=1$ boundary without radiation pressure from Fig~\ref{fig:metric_limb_prediction_generic_1e-5mbar_1000K}. We note that the two boundaries correspond to two distinct physical frameworks and only illustrate how the inclusion of radiation pressure modifies the particle size. Details of the calculation of $Q_\text{pr}$, where we present a smooth broken power-law fit to the Socrates-computed $Q_\text{pr}$ of soot-like haze so as to maximise computational flexibility, can be found in Appendix~\ref{sec:appendix_Qpr_of_soot}. As discussed in Sec.~\ref{subsec:analytical_theory}, the additional impact from radiation pressure acts to strengthen gravitational settling. As a result, Fig.~\ref{fig:metric_limb_prediction_generic_1e-5mbar_1000K_radiation_pressure} shows a decrease of the estimated maximum radius when compared to Fig.~\ref{fig:metric_limb_prediction_generic_1e-5mbar_1000K}. 

\begin{figure}
\centering
 \includegraphics[width=80mm]{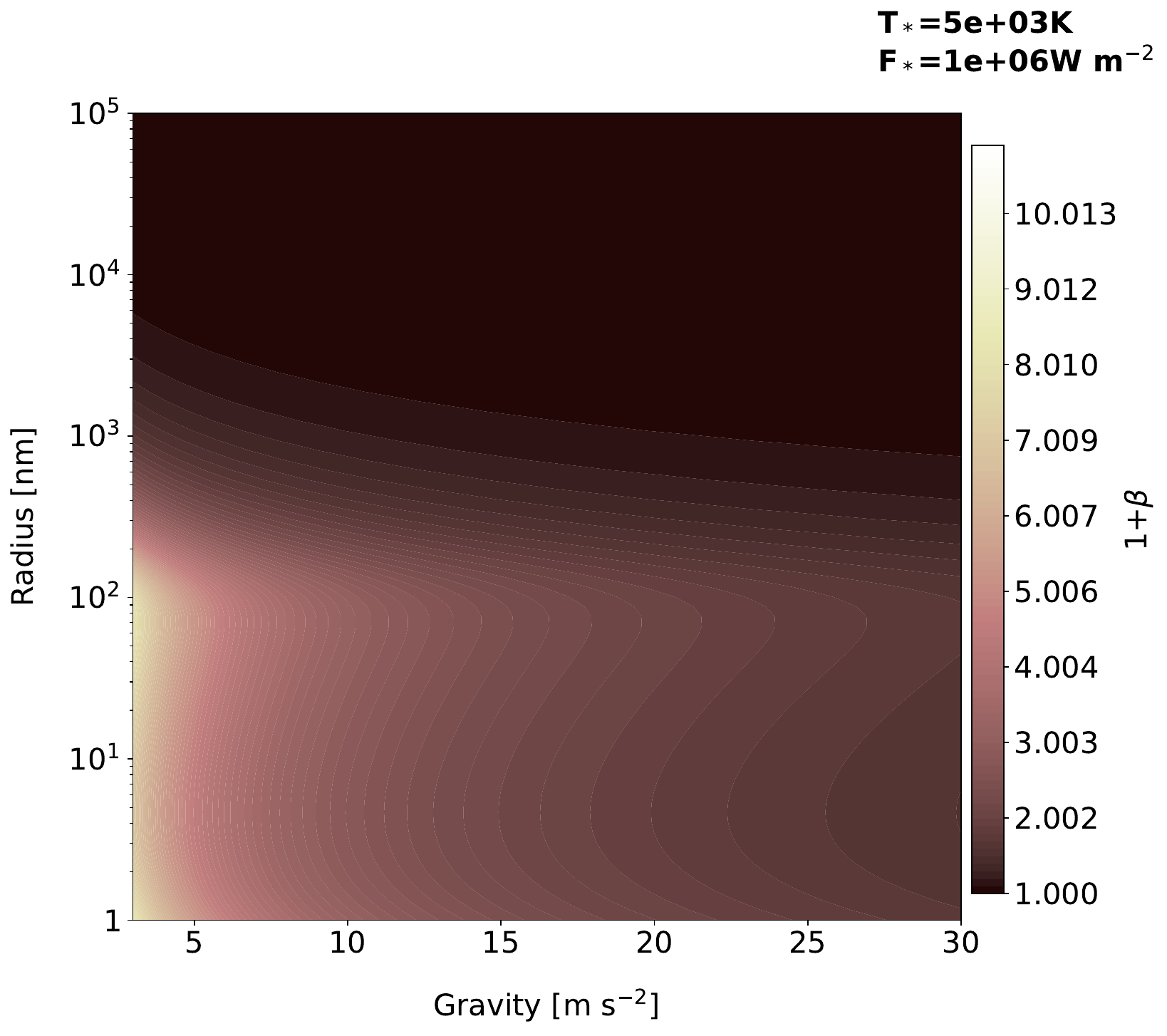}
 \caption{$1+\beta_{\text{rad}}$ computed at different particle radii and surface gravities, assuming $T_\ast=5000$\,K and $F_\ast=1\times10^6$\,W\,m$^{-2}$.}
 \label{fig:1_beta_F1e6}
\end{figure}

Fig.~\ref{fig:metric_limb_prediction_generic_1e-5mbar_1000K_radiation_pressure} also shows that the estimated maximum particle size for the HD\,209458b case decreases the most under the influence of radiation pressure, compared to HD\,189733b and WASP-39b. The full comparison is summarised in Tab.~\ref{tab:MAD_comparison_RP}. This difference can be understood from the dependence of $1+\beta_\text{rad}$ on particle radius and surface gravity from Equation~(\ref{eqn:beta_rad}). Fig.~\ref{fig:1_beta_F1e6} shows the value of $1+\beta_\text{rad}$ as a function of radius and gravity, assuming $T_\ast=5000$\,K and $F_\ast=1\times10^6$\,W\,m$^{-2}$. Fig.~\ref{fig:1_beta_F1e6} shows that in general $1+\beta_\text{rad}$ decreases as gravity increases and/or radius increases. Yet, there are two peaks at radius of $\sim$1\,nm and $\sim$80\,nm. The peak which occurs at radius of $\sim$1\,nm (smallest value in our parameter space) is due to the fact that $\beta_\text{rad}$ is inversely proportional to radius (Equation~(\ref{eqn:beta_rad})). Therefore the smaller the particle, the larger $\beta_\text{rad}$. Another peak which occurs at a radius of $\sim$80\,nm is due to the fact that $Q_\text{pr}$ (Equation~(\ref{eqn:Qpr})) peaks at $\sim$200\,nm (see Appendix~\ref{sec:appendix_Qpr_of_soot} for detailed calculations). In other words, $1+\beta_\text{rad}$ exhibits two maxima as $\beta_\text{rad}$ depends on both radius and $Q_\text{pr}$, where the latter is also a function of radius but in a separate form (see Equation~(\ref{eqn:Qpr}) and Appendix~\ref{sec:appendix_Qpr_of_soot}). Similar behaviour of $Q_\text{pr}$ is also found in \citet{Owen_etal2025}, who reported a peak between $\sim$200--300\,nm. However, their analysis only extended down to particle radius of 10\,nm and therefore do not capture the second maximum at smaller particle size identified in this work. For HD\,209458b the particle radius that balances horizontal advection and gravitational settling lies near the maximum at larger particle size. As a result, radiation pressure is especially effective and the maximum particle size is reduced the most compared to the other planets we consider. In short, the predicted maximum particle depends jointly on the effectiveness of advection relative to settling, and its susceptibility to radiation pressure. Changing either factor can shift the predicted particle radius.


\begin{table}
\centering
	\caption{Comparison of the estimated maximum radius [nm] for a particle to be transported from the evening limb to the morning limb without radiation pressure (noRP) and with radiation pressure (withRP).}
	\label{tab:MAD_comparison_RP}
        \begin{tabular}{ |l |l|l|l| }
        \hline
        Planet & noRP [nm] & withRP [nm] & $\frac{\text{(noRP)} - \text{withRP}}{\text{(noRP)}} [\%]$ \\
        \hline
        HD\,189733b & 58.1 & 29.9 & 48.5\\
        \hline
        HD\,209458b & 310 & 95.6 & 69.2\\
        \hline
        WASP-39b & 1450 & 945 & 34.8\\
		\hline
	\end{tabular}
\end{table}

\section{Discussion}
\label{sec:discussions}

In this work, we present a novel analytical framework that estimates how planetary radius, surface gravity, and the size of haze particles impact both the horizontal and vertical haze transport timescales. To do so, we incorporate the effect of atmospheric circulation, gravitational settling and radiation pressure in an analytic scaling theory. We introduce $\Psi_\text{HALD}$ (Haze Asymmetric Limb Distribution) which is the ratio between the downward and horizontal transport timescale ($\tau_v/\tau_a$). The threshold of $\Psi_\text{HALD}=$1 provides an estimate of the maximum particle radius that haze can attain before settling completely dominates, preventing the haze particles from reaching the morning limb and trapped within the nightside gyres and resulting in a higher haze mass mixing ratio over the morning limb than the evening limb. In other words, with the underlying assumption that haze particles would subsequently be trapped within the nightside gyres once reaching the morning limb, this HALD framework predicts an estimated size such that particles larger than it cannot maintain a higher or comparable concentration over the morning limb than the evening limb, and particles smaller than the predicted size can remain abundant on the morning limb and/or the evening limb, depending on the details of the atmospheric circulation such as local upwelling and downwelling. 

The results estimated by this framework agree well with previous 3D GCM studies of haze distributions, where different particle radii and surface gravities where explored (see Tab~\ref{tab:MAD_comparison}). We found that, on low-gravity planets, large particles can be advected efficiently towards the morning limb. Whereas on high-gravity planets, only small particles can be efficiently advected, limiting the maximum size distribution found over the morning limb. In both regimes, if haze is present in the atmosphere, at least a fraction of it is likely to be trapped around the nightside gyres instead of completely settling out over the nightside, as shown in our simulations (Fig.~\ref{fig:f1e-13_MMR_1e-5bar_MAD}), \Mtwentyfive\ and \citet{Steinrueck_etal2021}. At deeper pressures as the settling velocity decreases and the transport is dominated by the super-rotating jet (Fig.~\ref{fig:tw_ts_ta_MAD}), similar to what is found in \citet{Tsai_etal2024} and \citet{Owen_etal2025}, haze particles can be found in the morning limb more easily.

\subsection{Complexity of Atmospheric Circulation}
\label{subsec:complexity_of_atmospheric_circulation}

We note, however, that this framework provides only a first-order estimate of the limb haze distribution as it reduces the complex circulation to two generalised directions of motion, and adopts a single set of atmospheric parameters across different hot-Jupiters which could yield a biased representation of the atmospheric dynamics. As such, it does not capture how individual components of the circulation, and the localised advection patterns, shape the limb haze structure. Differences in equilibrium temperature, metallicity and composition could produce distinct thermal structures and winds, and thereby spatially varying horizontal and vertical timescales. The locations and relative strengths of upwelling and downwelling regions will also modulate how efficiently material is transported eastward, affecting the degree to which the atmosphere exhibits more haze over the morning or evening limb. Furthermore, the current HALD framework expects that haze particles reaching the morning limb may subsequently accumulate within the nightside gyres. However, the trapping process itself is not explicitly modelled. The amount of trapping depends on the detailed three-dimensional atmospheric circulation, including the location and morphology of the nightside gyres, as well as the residence timescale of particles within, all of which determine how effectively haze particles accumulate after reaching the nightside. In addition, the nightside gyres are primarily located at mid-latitudes rather than over the entirety of morning limb. Consequently, only particles transported into these regions are expected to experience efficient trapping, while particles transported along other latitudes may continue to advect around the planet. These processes are not captured by the present analytical framework and may therefore influence the amount of haze accumulated over the morning limb and therefore the extent of asymmetry in haze distribution. Future work is needed to investigate the details of the dynamics governing particle trapping within the nightside gyres.

Additionally, our current framework parametrises the resolved large-scale vertical transport using a characteristic hemispherically-averaged downward velocity. An alternative description of the same large-scale circulation is through an effective vertical mixing process, which can be parameterised as a diffusive process using an eddy diffusion coefficient $K_{zz}$. This would yield a characteristic timescale $\tau_{\text{mix}} = H^2/K_{zz}$ . In this work, we adopt the transport formulation as our analytical framework is aiming to estimate the competition between hemispheric horizontal transport and the net vertical removal of haze particles. The relationship between this transport timescale and an equivalent $K_{zz}$ requires a further in-depth analysis which is beyond the scope of this work. Moreover, transport of particles also occurs through sub-grid processes, such as small-scale eddy driven mixing.  Directly comparing haze transport across different model formulations is therefore challenging due to the range of scales included in the dynamical mixing, and the lack of a detailed understanding of the contribution of sub-grid mixing. Performing high-resolution simulations can be illustrative. However, if key physical processes are omitted, increased resolution alone might not improve accuracy of model predictions.

In our case, since we neglect small-scale eddies and diffusive processes in our large-scale framework, the mixing may be underestimated. For example, as the production of hazes generally peaks at lower pressures (where stellar flux is large), and removal of hazes occurs at deeper pressures, this results in a positive concentration gradient with respect to height. In such a situation, these small-scales eddies and diffusive processes would transport haze downward along the concentration gradient, effectively acting as a sink for the observable atmosphere. Since both large-scale downwelling and small-scale vertical mixing can help remove haze at lower pressures, our inclusion of only the large-scale transport would impact the inferred gravitational settling strength required to maintain haze particles at different pressure levels. A combined approach of using observationally inferred particle sizes at various pressures together with model predictions can be used to explore the strength and behaviour, across planetary parameters, of the combined mixing. 


\begin{figure}
\centering
 \includegraphics[width=80mm]{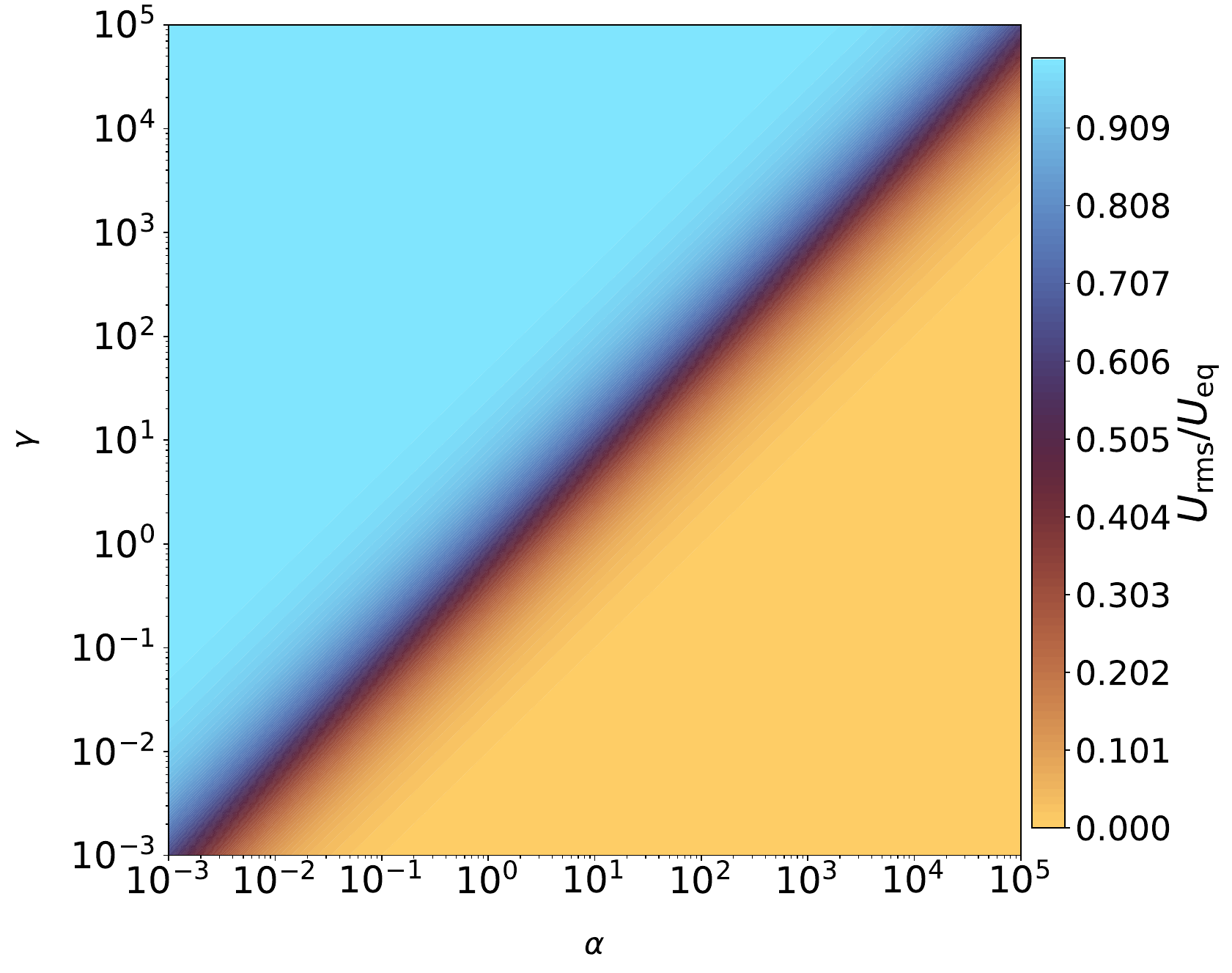}
 \caption{$U_\text{rms}/U_\text{eq}$ from Equation~(\ref{eqn:U_rms}) as a function of $\alpha$ and $\gamma$.}
 \label{fig:Urms_over_Ueq}
\end{figure}
Furthermore, changes to the assumptions made in our analytical framework would also affect our results. For instance, in a non-isothermal atmosphere where temperature decreases with pressure, $\tau_\text{wave}$ would reduce (Equation~(\ref{eqn:tau_wave})). In addition, the pressure level corresponding to the observable photosphere $p_{\tau=1}$ would vary with for example gaseous compositions and aerosols. Changes in $p_{\tau=1}$ would modify $\Delta \ln p$ and subsequently $U_\text{eq}$ (Equation~(\ref{eqn:U_eq})) and $\tau_\text{rad}$ (Equation~(\ref{eqn:tau_rad})). For example, if transmission is probed at pressures <10\,mbar, the value of $\Delta \ln p$ will increase. An increase of the value of $\Delta \ln p$ would act to lower the values of $\alpha$ and $\gamma$ (Equation~(\ref{eqn:U_eq})). Similarly, in a dense atmosphere or with a low heat capacity where $\tau_\text{rad}$ decreases (Equation~(\ref{eqn:tau_rad})), the values of $\alpha$ and $\gamma$ would increase. Fig.~\ref{fig:Urms_over_Ueq} demonstrates how changing $\alpha$ and $\gamma$ would impact $U_\text{rms}$, and it shows that decreasing $\alpha$, and/or increasing $\gamma$ would increase $U_\text{rms}$, and vice versa. Last but not least, if an atmosphere has a larger day-night temperature difference in radiative equilibrium $\Delta T_\text{eq}$, $U_\text{eq}$ and $U_\text{rms}$ will increase (Equations~(\ref{eqn:U_eq}) and~(\ref{eqn:U_rms})). Increasing $U_\text{rms}$ will also alter $W$ through Equation~(\ref{eqn:W}), changing the maximum particle size estimated by the HALD framework based on two particle size regimes: 1) For small particles where the resulting downward motion is predominantly attributed to vertical advection due to weak gravitational settling ($V_s << W$), increasing $U_\text{rms}$ will increase $W$ as $W \propto U_\text{rms}^2$. Since $\Psi_\text{HALD}\propto U_\text{rms}/W \propto 1/U_\text{rms}$, increasing $U_\text{rms}$ reduces $\Psi_\text{HALD}$ eventually, resulting in an overall stronger settling motion than horizontal advection. 2) For large particles where the resulting downward motion is predominantly attributed to gravitational settling than vertical advection ($V_s >> W$), since $\Psi_\text{HALD}\propto U_\text{rms}/V_s$, increasing $U_\text{rms}$ increases $\Psi_\text{HALD}$ eventually, meaning larger particles can be advected efficiently to the morning limb. In other words, changes to the assumption made in Sec.~\ref{subsec:analytical_theory}, depending on the particle size regime, will alter the estimated maximum particle size that can be advected efficiently to the morning limb. 

Considering the assumptions used to construct the analytical model, this HALD framework will be the most accurate in the settling-dominated regime, where the resulting haze distribution is controlled primarily by strong gravitational settling. As shown in Sec.~\ref{subsec:advection_vs_settling}, the particle radius estimated by the HALD framework is in general consistent with those diagnosed via 3D GCM simulations that explore large particle sizes (Fig.~\ref{fig:g_4_varying_r}) or a high-gravity planet (Fig.~\ref{fig:g_20_varying_r}). In contrast, for low-gravity planets and/or small particle sizes, the transport of haze particles becomes increasingly sensitive to the detailed circulation structure that is not captured by the analytical model. Ongoing work is being conducted which investigates the mechanisms that drive the deviation between the approximated maximum radius using our framework and the radius predicted by the GCM simulations (Mak et al. in prep). These dynamical approximations can also be improved in future work by performing detailed planet-specific 3D simulations, especially in the scenario where $\Psi_\text{HALD}>1$, which will resolve the detailed advective patterns, eddy fluxes, and vertical structure that control limb-to-limb differences in haze distribution.

\subsection{Microphysical and Radiative Properties of Haze}
\label{subsec:microphysical_and_radiative_properties_of_haze}

Another uncertainty arises from our neglect of haze coagulation, which would depend on for example the air and particle density, monomer size, coagulation kernel, sticking efficiency and surface energy of haze particles \citep{Lavvas_and_Koskinen_2017}. If the coagulation timescale is shorter than the local transport or removal timescale, haze particles may grow during transport through the atmosphere. As a result, the balance between horizontal transport and vertical removal would evolve as particles move through the atmosphere, potentially growing in size which results in an increasing strength of gravitational settling. Future microphysical studies are needed to self-consistently couple haze microphysics to atmospheric transport. Additionally, the radiative feedback from haze would further impact the circulation, particularly for large values of $F_0$ that could result a high haze concentration within the atmosphere. Such feedback may alter the strength and location of the super-rotating jet or the position and morphology of nightside gyres (\Mtwentyfive). This HALD framework would therefore produce the closest agreement analytically with the GCM simulations under a low $F_0$, or when the atmosphere consists of a haze type that has very weak absorption strength. However, \Mtwentyfive\ demonstrate that the mechanisms controlling the global haze distribution remains similar regardless of the strength of the haze radiative forcing, suggesting that our basic predictive framework should be robust. More realistically, whether or not induced by radiation pressure \citep{Owen_etal2025}, there is likely to be both eastward and westward transport of haze that will effectively compete. The haze that is deposited over the morning limb is set by a combination of eastward transport from the night side and westward transport from the day side. Future work diagnosing tracer mixing of haze in 3D GCMs is required to determine the relative influence of each flow on the morning limb haze distribution.

\subsection{Complexity of Radiation Pressure}
\label{subsec:complexity_of_radiation_pressure}

This work has assumed fixed stellar parameters (a Sun-like star) and a single representative haze type across hot-Jupiters, implying that additional work is required to fully diagnose the role of radiation pressure. For instance, a larger $F_\ast$ would increase $\beta_\text{rad}$ (Equation~(\ref{eqn:beta_rad})), hence strengthening the effect of radiation pressure and lowering the estimated particle size for a more haze build-up over the morning limb. A different stellar type with different $T_\ast$ would also produce different $Q_\text{pr}$ \citep{Pawellek_etal2019}. Soot, which has been commonly used in the modelling of haze in the atmospheres of hot-Jupiters, has a very strong $Q_\text{pr}$ \citep{Owen_etal2025}. The settling strength might be overestimated if a haze type with a much weaker extinction property is considered. We therefore expect that varying stellar properties and haze compositions will primarily rescale the strength of radiation pressure, and thus shift the estimated particle size, without qualitatively altering the trend shown in Fig.~\ref{fig:metric_limb_prediction_generic_1e-5mbar_1000K_radiation_pressure}. Moreover, the simplified treatment of radiation pressure introduces further uncertainties as we neglect the decrease of diffused shortwave flux in deeper pressures \citep{Pierrehumbert_2010}, the reduction in the vertical component of the force due to the non-zero incidence angle of stellar radiation, and the absence of radiation pressure on the nightside. These effects would weaken the influence of radiation pressure, implying that the settling strength is overestimated in our calculations. As a result, the true radius is expected to lie between the values predicted by the no-radiation-pressure and the full-radiation-pressure frameworks. However, despite of the simplifying assumptions made in this work as discussed in this Section, by adopting the same calculations presented in Sec.~\ref{subsec:analytical_theory} with the appropriate parameters for a specific system, one can compute $\Psi_\text{HALD}$ for the target planet, enabling more detailed and efficient planning of observational campaigns. 

\section{Conclusions}
\label{sec:conclusions}

This work presents a novel framework to estimate the maximum radius that a photochemical haze particle can have in order to be advected to the morning limb efficiently. Assuming efficient trapping of haze particles within the nightside gyres, this corresponds to the maximum particle size capable of producing a higher or comparable haze concentration over the morning limb relative to the evening limb. In this work, we develop an analytical model, which takes into account the effect of particle radius, surface gravity, and radiation pressure and provide a rapid first-order approximation of the limb haze distribution. It can be directly connected to limb asymmetries observed by JWST, as well as to guide future observational campaigns, making it a useful tool for planning target selection and for understanding the potential limb asymmetries in transmission spectra prior to investing in computationally expensive 3D simulations. If a larger transit depth of a morning limb is revealed in transit limb observations, this would indicate a size-limited distribution for photochemically generated hazes in which only certain particle sizes are efficiently transported to the morning limb, constraining haze microphysics, transport, the atmospheric dynamics and vertical mixing within the atmosphere. The estimated particle sizes at the pressure level probed by transmission spectra could be compared with our analytical framework to quickly assess whether photochemically produced hazes are capable of driving the observed asymmetries. These inferences would also be highly valuable for benchmarking 3D GCMs and facilitating our understanding of the atmospheric dynamics of hot-Jupiters. Even if future observations do not show a larger morning limb but find a super-Rayleigh slope indicative of potential haze \citep{Ohno_and_Kawashima_2020}, whether at the evening limb or full transmission spectra, this framework could be used to diagnose why haze is not observed over the morning limb. Possible explanations include small particle sizes and/or low production rates, both of which would result in weak extinction effect, or that the haze is removed by unknown processes or masked by clouds. 


\section{Acknowledgments}
We thank the anonymous reviewer for their constructive feedback, which have improved the quality of this work. We acknowledge funding from the Croucher Postdoctoral Fellowship, funded by the Croucher Foundation, which made this work possible. We thank Krisztian Kohary for maintaining the UM. This work was supported by a UKRI Future Leaders Fellowship and extension [grant numbers MR/T040866/1 \& MR/Z000122/1], a Science and Technology Facilities Council Consolidated Grant [ST/R000395/1] and the Leverhulme Trust through a research project grant [RPG-2020-82]. Material produced using Met Office Software. We acknowledge use of the Monsoon3 system, a collaborative facility supplied under the Joint Weather and Climate Research Programme, a strategic partnership between the Met Office and the Natural Environment Research Council. This work used the DiRAC Complexity system, operated by the University of Leicester IT Services, which forms part of the STFC DiRAC HPC Facility (www.dirac.ac.uk). This equipment is funded by BIS National E-Infrastructure capital grant ST/K000373/1 and STFC DiRAC Operations grant ST/K0003259/1. DiRAC is part of the National e-Infrastructure. 

\section*{Data Availability}
The simulation data presented in this study will be shared on reasonable request to the corresponding author.


\bibliographystyle{mnras}
\bibliography{reference} 


\appendix
\section{Q$_{pr}$ of Soot}
\label{sec:appendix_Qpr_of_soot}
\begin{figure}
\centering
 \includegraphics[width=70mm]{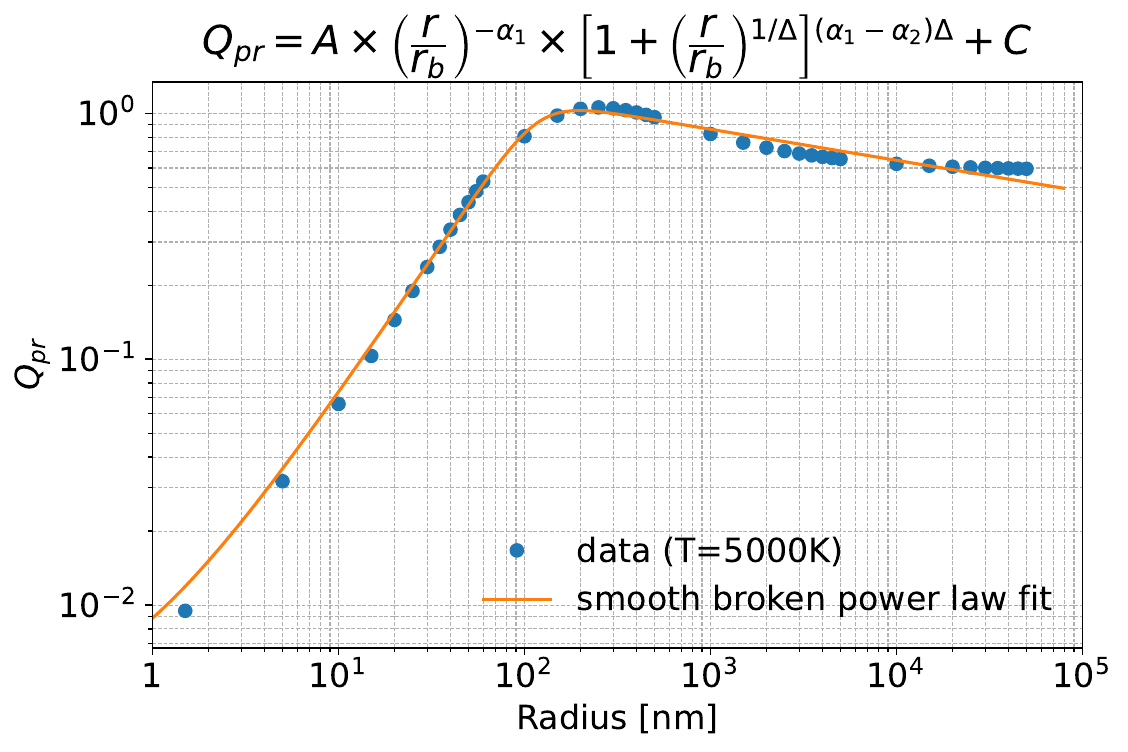}
 \caption{$Q_{\text{pr}}$ computed at different radii (circles), assuming a stellar temperature $T_\ast=5000$\,K, using $Q_{\text{ext},\lambda}$, $Q_{\text{sca},\lambda}$ and $\langle g_\lambda (r)\rangle$ calculated from Socrates (Sec.~\ref{subsubsec:haze_model}). A smooth broken power-law fit is applied to the Socrates-computed data points.}
 \label{fig:Qpr_soot_T5000_fit}
\end{figure}

As given in Equation~(\ref{eqn:Qpr}), $Q_{\text{pr}}$ depends on the optical profile of the particle. Fig.~\ref{fig:Qpr_soot_T5000_fit} shows the computed $Q_{\text{pr}}$ for soot-like haze at different radii, assuming a Sun-like star with stellar temperature $T_\ast=5000$\,K, using $Q_{\text{ext},\lambda}$, $Q_{\text{sca},\lambda}$ and $\langle g_\lambda (r)\rangle$ calculated from Socrates (Sec.~\ref{subsubsec:haze_model}). Fig.~\ref{fig:Qpr_soot_T5000_fit} shows that $Q_{\text{pr}}$ reaches a maximum at radius $\sim$200\,nm, and decreases slightly before reaching a plateau at radius $\geq$200\,nm. $Q_{\text{pr}}$ drops off significantly at radius $\leq$200\,nm. For efficient calculation of $Q_{\text{pr}}$ at any given radius performed within this study, and without involving Socrates, a smooth broken power-law fit is applied to the already-computed $Q_{\text{pr}}$, described as 
\begin{equation}
\label{eqn:smooth_broken_power_law_fit}
Q_{\text{pr}} = A \left( \frac{r}{r_b} \right)^{-\alpha_1} \left[ 1 + \left( \frac{r}{r_b} \right)^{1/\Delta}\right]^{(\alpha_1 - \alpha_2)\Delta} + C \quad,
\end{equation}
where $r$ is the radius and the rest are constants. The fit is performed using the \texttt{curve\_fit} function in the \textsc{SciPy} library \citep{Virtanen_etal2020_scipy}, yielding $A=$ 1.1253, $r_b=1.1994\times10^{-7}$, $\alpha_1$= -1.1165, $\alpha_2=$ 0.1272, $\Delta=$ 0.2287 and $C=$ 0.003472.


\bsp	
\label{lastpage}
\end{document}